\documentclass[twoside,a4paper,fleqn,titlepage,preprintnumbers,nofootinbib,%
               prstab,12pt,showpacs,showkeys]{revtex4-1}
\usepackage{url}                
\usepackage{graphicx}
\usepackage{amssymb}
\usepackage{epstopdf}
\usepackage{subfigure}
\begin{document}

\title{A Toroidal Magnetised Iron Neutrino Detector (MIND) for a Neutrino Factory}
\author{A.~Bross}
\author{R.~Wands}
\affiliation{Fermi National Accelerator Laboratory,  Batavia, IL, USA.}\author{R.~Bayes}
\author{A.~Laing}
\author{F.J.P.~Soler}
\affiliation{School of Physics \& Astronomy, University of Glasgow,
  Glasgow, UK.}
\author{A. Cervera Villanueva}
\author{T.~Ghosh}
\author{J.J.~G\'omez Cadenas}
\author{P.~Hern\'andez}
\author{J.~Mart\'in-Albo}
\affiliation{IFIC, CSIC \& Universidad de Valencia, Valencia, Spain.}
\author{J.~Burguet-Castell}
\affiliation{Universitat de les Illes Balears, Spain.}
\begin{abstract}
  A neutrino factory has unparalleled physics reach for the discovery
  and measurement of CP violation in the neutrino sector. A far
  detector for a neutrino factory must have good charge identification
  with excellent background rejection and a large mass. An elegant
  solution is to construct a magnetized iron neutrino detector (MIND)
  along the lines of MINOS, where iron plates provide a toroidal
  magnetic field and scintillator planes provide 3D space points. In
  this report, the current status of a simulation of a toroidal MIND
  for a neutrino factory is discussed in light of the recent
  measurements of large $\theta_{13}$. The response and performance
  using the 10~GeV neutrino factory configuration are presented. It is
  shown that this setup has equivalent $\delta_{CP}$ reach to a MIND
  with a dipole field and is sensitive to the discovery of CP
  violation over 85\% of the values of $\delta_{CP}$.
\end{abstract}

\date{\today}                                           

\maketitle

\section{Introduction}
The neutrino factory is a new type of accelerator facility in which a
neutrino beam is created from the decay of muons in flight in a
storage ring. This facility can be used to study neutrino oscillations in a variety of
oscillation channels \cite{Geer:1997iz} and can be used to determine the neutrino mass hierarchy, whether the mass squared difference between neutrino mass eigenstates 
$\Delta m^2_{13}$ is positive or negative (inverted or normal mass hierarchy), and CP 
violation in the neutrino sector. The $\nu_e \rightarrow \nu_\mu$
oscillation \cite{DeRujula:1998hd}, identified through the so-called ``golden channel'' 
in which the charged current interactions of the $\nu_\mu$ produce muons of the opposite charge to those stored in the storage ring (wrong-sign muons) \cite{Cervera:2000kp},
is the most promising channel to explore CP violation at a neutrino factory. The physics capabilities and the design of the neutrino factory is carried out as part of the International Design Study for a Neutrino Factory (IDS-NF) \cite{IDS-NF}, partially funded through the EUROnu project \cite{EUROnu}.

In this paper we will describe the requirements and design of a neutrino factory far detector
and the analysis carried out to extract the wrong-sign muon neutrino oscillation signal. The far detector at a neutrino factory  \cite{Choubey:2011zz} requires
excellent reconstruction and charge detection efficiency. These
capabilities are best encompassed using a large magnetized iron
neutrino detector (MIND). For the discussion below, a MIND design with
a toroidal magnetic field, based on experience from the MINOS far
detector \cite{Michael:2008bc} is described with detailed simulations.

With the measurement of large $\theta_{13}$
\cite{Abe:2011sj,An:2012eh,Ahn:2012nd,Abe:2011fz,Adamson:2011qu} the
physics goals of the neutrino factory are focused on the measurement
of CP violation and the mass hierarchy in neutrino oscillations. This
requires re-optimization of the neutrino factory baseline and
re-evaluating the detector for this physics goal. The new experimental
setup consists of a single 2000~km baseline from a muon storage ring
wherein both $\mu^+$ and $\mu^-$ decay at energies of 10~GeV to a
single 100 kTon detector with a toroidal magnetic field. The analysis
described here improves and simplifies a previous analysis based on a
100 kton detector with a dipole magnetic field \cite{Bayes:2012ex}.

The neutrino beam from a neutrino factory contains both
$\nu_{\mu}(\bar{\nu}_{\mu})$ and $\bar{\nu}_{e}(\nu_{e})$ resulting
from the decay of $\mu^-(\mu^+)$ in a storage ring. As such, there are
a number of possible oscillation channels, summarized in
Table~\ref{tab:osc}. The MIND is optimized to exploit the golden
channel oscillation as this has an easily identified signal; a muon
with a sign opposite to that in the muon storage ring.  With the
exception of the silver channel oscillation, which re-enforces the
golden channel signal, other oscillation channels are treated as
background. The event selection required to produce the best signal
and background rates for the measurement of CP violation is the focus
of ongoing optimization.

This paper describes the detector design (Section~\ref{sec:design})
before discussing the simulation (Section~\ref{sec:sim}) and event
reconstruction (Section~\ref{sec:rec}). Event selection is given in
Section~\ref{sec:anal} and the resulting efficiency and background
rates are presented in Section~\ref{sec:Eff}. Finally, the sensitivity
and precision of the detector to CP violation is presented in
Section~\ref{sec:sense}.

\begin{table}[htdp]
\caption{List of oscillations expected at a neutrino factory.}
\begin{center}
\begin{tabular}{|l|c|c|}
\hline\hline
& Store $\mu^{+}$  &Store $\mu^{-}$ \\
\hline
Golden Channel & $\nu_{e}\to\nu_{\mu}$ & $\bar{\nu}_{e}\to\bar{\nu}_{\mu}$ \\
$\nu_e$ Disappearance Channel & $\nu_{e}\to\nu_{e}$ & $\bar{\nu}_{e}\to\bar{\nu}_{e}$ \\
Silver Channel & $\nu_{e}\to\nu_{\tau}$ & $\bar{\nu}_{e}\to\bar{\nu}_{\tau}$ \\
\hline
Platinum Channel & $\bar{\nu}_{\mu}\to\bar{\nu}_{e}$ & $\nu_{\mu}\to\nu_{e}$ \\
$\nu_{\mu}$ Disappearance Channel & $\bar{\nu}_{\mu}\to\bar{\nu}_{\mu}$ & $\nu_{\mu}\to\nu_{\mu}$ \\
Dominant Oscillation & $\bar{\nu}_{\mu}\to\bar{\nu}_{\tau}$ & $\nu_{\mu}\to\nu_{\tau}$ \\
\hline\hline 
\end{tabular}
\end{center}
\label{tab:osc}
\end{table}%

\section{Detector Design}\label{sec:design}

\begin{figure}[htbp]
\begin{center}
\subfigure[Engineering drawing of the MIND plate]{
	\includegraphics[width=0.475\textwidth]{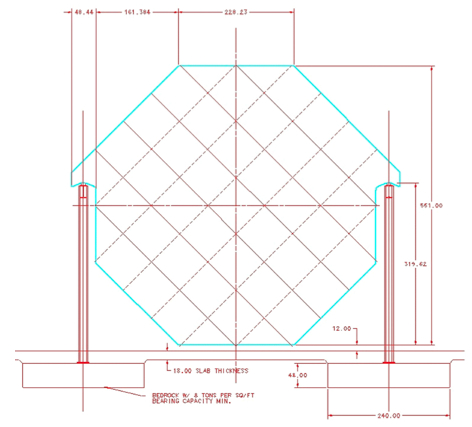}
	\label{fig:plane}
}
\subfigure[Orthographic view of MIND]{
	\includegraphics[width=0.475\textwidth]{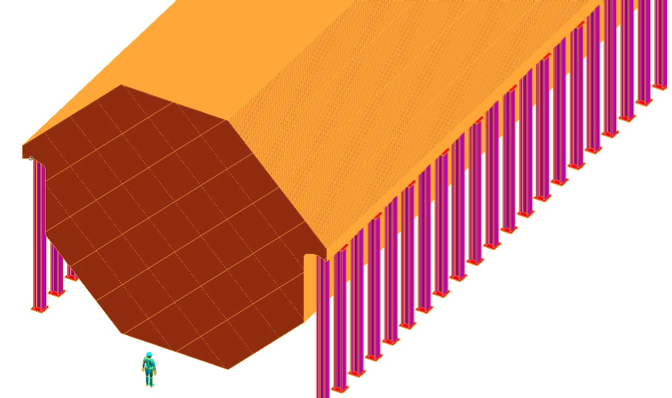}
	\label{fig:ortho}
}
\caption{Schematic representations of the MIND for a neutrino factory.}
\label{fig:det}
\end{center}
\end{figure}

The MIND is an iron-scintillator calorimeter with an octagonal cross
section 14~m high and 14~m in width (Fig. \ref{fig:det}). Modules of
3~cm thick iron plates and a 2~cm thick lattice of scintillating bars
compose the 100 kTon bulk of the detector. The iron planes provide the
structural strength for the calorimeter as well as the magnetic field
necessary for charge discrimination. Due to practical constraints, the
iron planes are to be constructed of strips of steel 1.5~cm thick and
2~m wide. By arranging these strips in a lattice configuration, the
resulting structure possesses the necessary rigidity and tensile
strength to support its own weight by two ``ears" projecting from the
sides of the plate, with distortions in the plate dimensions of less
than 2~mm.

\begin{figure}[htbp]
\begin{center}
\includegraphics[width=0.6\textwidth]{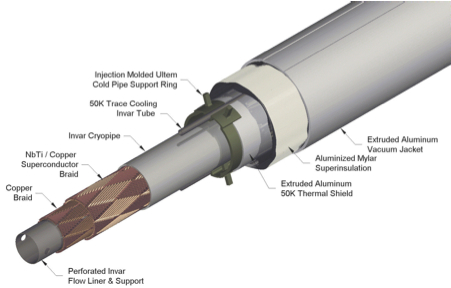}
\caption{The super-conducting transmission line, proposed as a current source for MIND}
\label{fig:STL}
\end{center}
\end{figure}

\begin{figure}[htbp]
\begin{center}
\includegraphics[width=0.55\textwidth]{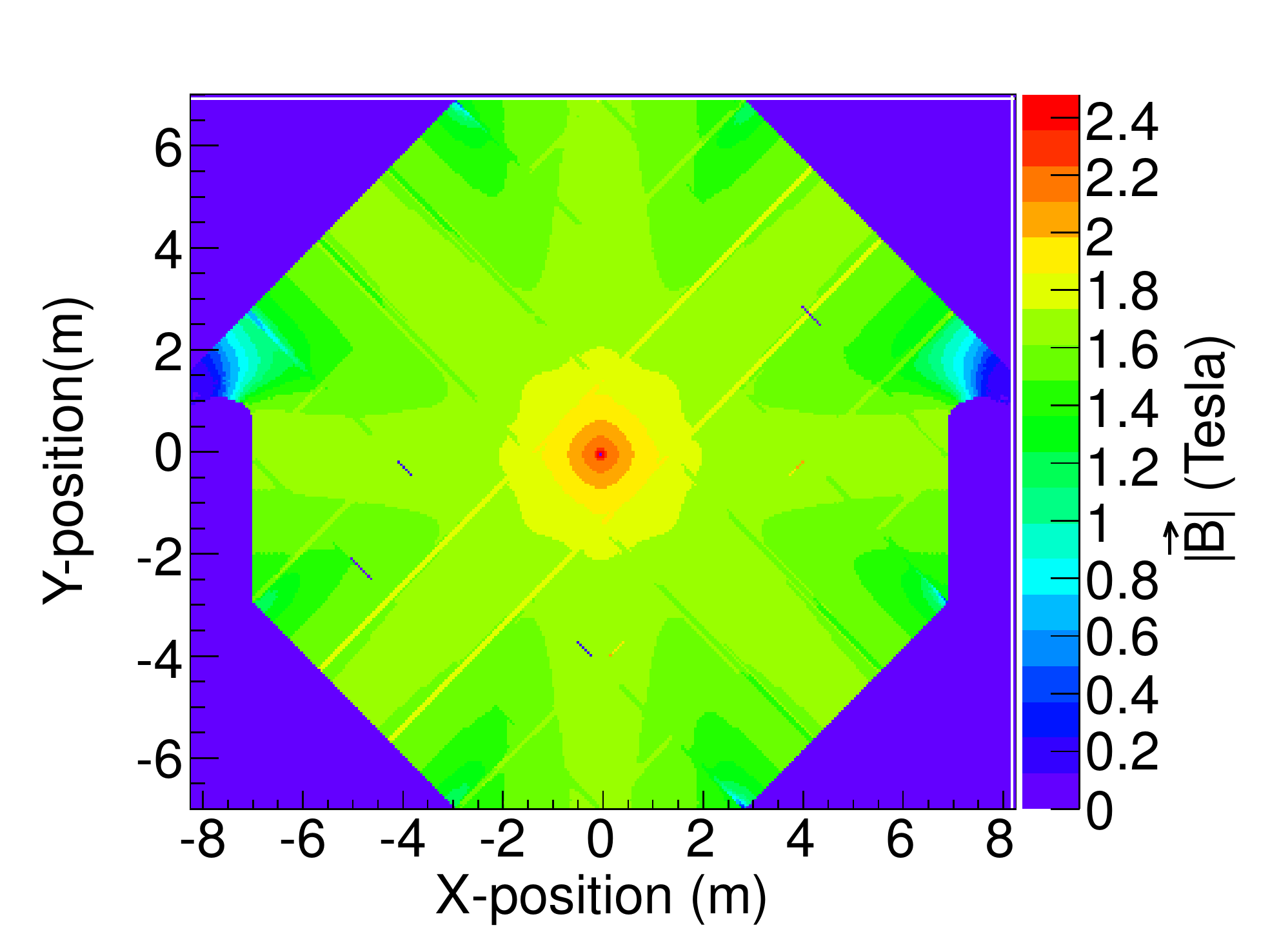}
\caption{The magnetic field in iron simulated from the finite element studies of the MIND plate assuming an STL current of 100~kA}
\label{fig:map}
\end{center}
\end{figure}

To induce the 1 Tesla magnetic field in the iron plate, a current of 100 kA through the centre of the detector is required. This current is to be carried by a super-conducting transmission line (STL), which consists of copper and copper/NbTi alloy braids contained by a cryogenic jacket, 7~cm in diameter \cite{Ambrosio:2001ej,828238}. The STL runs through a 10~cm bore along the central axis of the detector. A detailed diagram of the STL is shown in Fig.~\ref{fig:STL}. A map of the magnetic field in the iron has been generated using a finite element model of the detector plate. The simulated field map is shown in Fig.~\ref{fig:map}.

The detection of neutrino interactions is accomplished through the use
of scintillating bars arranged in a lattice to define a 3D space point
for the energy deposition of a passing particle. Assuming a coordinate
system for the detector such that the neutrino beam defines the
$z$-axis, perpendicular to the detector face, the scintillator bars
are arranged in a layer to measure the position of an event hit along
the $x$-axis and a layer to measure a hit position along the
$y$-axis. Each scintillator bar is rectangular with a
1~cm$\times$3.5~cm cross-section and spans the width of the
detector. A wavelength shifting fibre 1~mm thick runs down the centre
of the scintillating bar and is coupled at each end of the bar to a
silicon photomultiplier.

\section{Simulation}\label{sec:sim}

Neutrino interactions in the MIND simulation are generated using the
GENIE framework \cite{Andreopoulos:2009rq}. This framework reproduces
deep inelastic scattering (DIS), quasi-elastic scattering (QES),
single pion production, resonant pion production, coherent pion
production, and neutrino-electron elastic scattering
processes. Previous simulation studies for MIND \cite{Cervera:2010rz}
have been produced using LEPTO \cite{Ingelman:1996mq} and NUANCE
\cite{Casper:2002sd}. These packages are incomplete descriptions of
the neutrino interactions as they do not include such phenomena as
re-interaction within participant nuclei; an important feature in high
Z targets such as iron.

The detector geometry was constructed using the GEANT4 framework
\cite{1610988}. The geometry was defined with some flexibility in the
detector dimensions, including the transverse and longitudinal lengths
as well as the thickness of the iron and scintillator planes to allow
for optimization studies. The magnetic field, although basically
toroidal, is applied using a field map. Products of the neutrino
interaction events generated by GENIE are propagated through the
detector materials using the QGSP\_BERT physics list provided by
GEANT4.

\section{Reconstruction}\label{sec:rec}

Following simulation, the events are digitized in a very simple
way. The position and energy deposition for a hit in a given
scintillator plane are clustered in a 3.5~cm$\times$3.5~cm unit,
called a voxel, which is defined by the expected positions of the
scintillator bars in the transverse plane. The energy deposition is
attenuated over the distance of the hit from the edge of the detector
assuming an attenuation length of 5~m. The digitized hits are passed
to a reconstruction module.

The purpose of the reconstruction is to identify and fit potential
muon tracks resulting from charge current neutrino interactions. The
reconstruction uses algorithms provided by the RecPack toolkit
\cite{CerveraVillanueva:2004kt}. The majority of tracks are identified
from the event using a Kalman filtering algorithm. First a prospective
track is identified by looking for the longest set of planes with a
single digitized hit. A guess for an initial angle and momentum is
generated from this information and used for an initial
fit. Additional hits are then filtered into the track by looking for
hits that produce the smallest local $\chi^2$ value in planes with
multiple hit occupancies. The subset of events that do not have a set
of single occupancy planes are subjected to a cellular automaton
algorithm \cite{Abt:2002ay} for the identification of tracks within
events where the muon track is not separated from the hadron
activity. In either case the longest track is selected as the muon
trajectory passed to the fitting algorithm.

\begin{figure}[htbp]
\begin{center}
\includegraphics[width=0.65\textwidth]{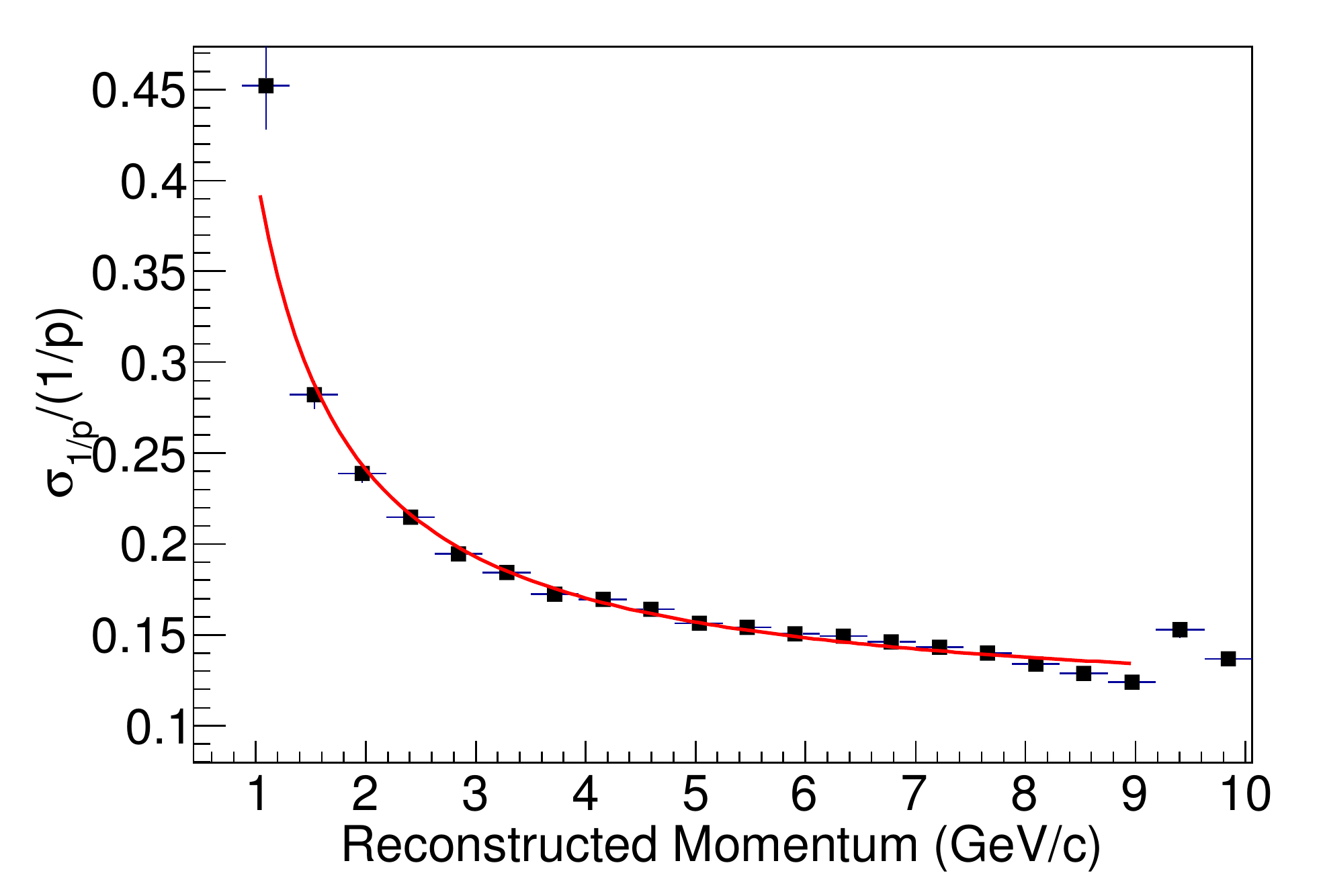}
\caption{Momentum resolution from the muon track reconstruction. The
  parametrization of the resolution measured from the simulation is
  shown with the red solid line.}
\label{fig:res}
\end{center}
\end{figure}

The identified muon tracks are subject to a Kalman fitting process to
determine their momentum and charge. The Kalman fitter uses a model to
predict the position from one hit to the next in a sequence correcting
for random noise, such as from multiple scattering, and allowing for
processes such as energy loss --- which is now included as a function
of momentum. An initial seed for the fit is determined from the
geometry of the muon track using the range of the muon track to supply
the momentum \cite{groom:2001ms} and the relative positions of the
beginning and end of the track in the bending plane to determine the
charge. This seed is passed to the fitting algorithm where the track
parameters are further refined. A successfully reconstructed track
survives the Kalman fitting process six times; twice during the track
identification stage where the track is fitted and filtered and four
more times during the fitting stage assuming different fitting
seeds. These algorithms are based on previous work
\cite{Cervera:2010rz,Bayes:2012ex}, but adapted to the new toroidal
magnetic field configuration. The momentum resolution resulting from
this fit is shown in Fig.~\ref{fig:res}. The behaviour of the
resolution on the inverse of the momentum ($1/p$) is parametrized as
follows;
\begin{equation}
\frac{\sigma_{1/p}}{1/p} = 0.24 - \frac{0.061}{p(GeV)} + 0.11 p.
\end{equation}

The neutrino energy is currently reconstructed using the combination
of the reconstructed muon momentum and the smeared true hadron energy,
$E_{had}$. This smearing assumes an energy resolution $\delta E_{had}$
measured from the MINOS CalDet test beam \cite{Michael:2008bc};
\begin{equation}
\frac{\delta E_{had}}{E_{had}} = \frac{0.55}{\sqrt{E_{had}}} \oplus 0.03.
\end{equation}
Since Ref. \cite{Michael:2008bc} does not provide the angular
resolution, this was taken from the measurements at the Monolith test
beam \cite{Bari:2003bt};
\begin{equation}
\delta \theta_{had} = \frac{10.4}{\sqrt{E_{had}}}\oplus \frac{10.1}{E_{had}}.
\end{equation}
Current work to explicitly identify and reconstruct the hadron showers
will remove the necessity of this smearing process for the generation
of a reconstructed energy. In the case of quasi-elastic (QES) events,
the neutrino energy is calculated from the expression
\begin{equation}
E_{\nu} = \frac{m_{N}E_{\mu} + \frac{1}{2}(m^{2}_{N'} - m^2_{\mu} - m^2_{N})}{m_{N} - E_{\mu} +|p_{\mu}|\cos\vartheta}
\end{equation}
where $\vartheta$ is the angle between the muon momentum vector and
the beam direction, $m_{N}$ is the mass of the initial state nucleon and $m_{N'}$ is
the mass of the final state nucleon in the processes $\nu_{\mu} + n
\to \mu^{-} + p$ and $\bar{\nu}_{\mu} + p \to \mu^{+} + n$.




\section{Analysis}\label{sec:anal}

Successfully reconstructed events are subjected to a series of cuts to
isolate the wrong sign muons resulting from $\nu_{e}\to\nu_{\mu}$
oscillations from backgrounds that are similar to neutral current
events. All cuts used in the analysis are summarized in
Table~\ref{tab:desc} and are similar to those from a previous analysis
\cite{Bayes:2012ex}. The first cut ensures that the event is
successfully reconstructed by the Kalman filter. The second cut
removes events for which the first scintillator hit appears less than
1.5~m from the end of the detector.  Tracks reconstructed with momenta
greater than 16~GeV are removed to reduce biases from non-physical
reconstructed neutrino energies. A cut is also applied requiring that
60\% of the candidate hits of the track are used in the final fit, to
avoid tracks with hard scattering events or other sources of noise.

\begin{table}
	\begin{tabular}{|r|l|}
	\hline\hline
	Event Cut & Description \\
	\hline
	Successful Reconstruction & Failed Kalman reconstruction of event removed \\
	Fiducial   				  & First hit of event is more than 1.5~m from end of detector \\
	Maximum Momentum 	  & Muon momentum less than 1.6$\times E_{\mu}$\\
	Fitted Proportion 		  & 60\% of track nodes used in final fit.\\
	Track Quality 			  & $\log(P(\sigma_{q/p}/(q/p) | CC) / P(\sigma_{q/p}/(q/p)|NC)) > -0.5$\\
	CC Selection 			  & $\log(P(N_{hit} | CC) / P(N_{hit}|NC)) > 1.0$\\
	Kinematic 			  & $Q_t > 0.15 GeV$ \\
	\hline\hline
	\end{tabular}
	\caption{Description of cuts used in the selection of good events from the simulation}
	\label{tab:desc}
\end{table}

\begin{figure}[htbp]
\begin{center}
\subfigure[Stored $\mu^+$ experiment]{
\includegraphics[width=0.475\textwidth]{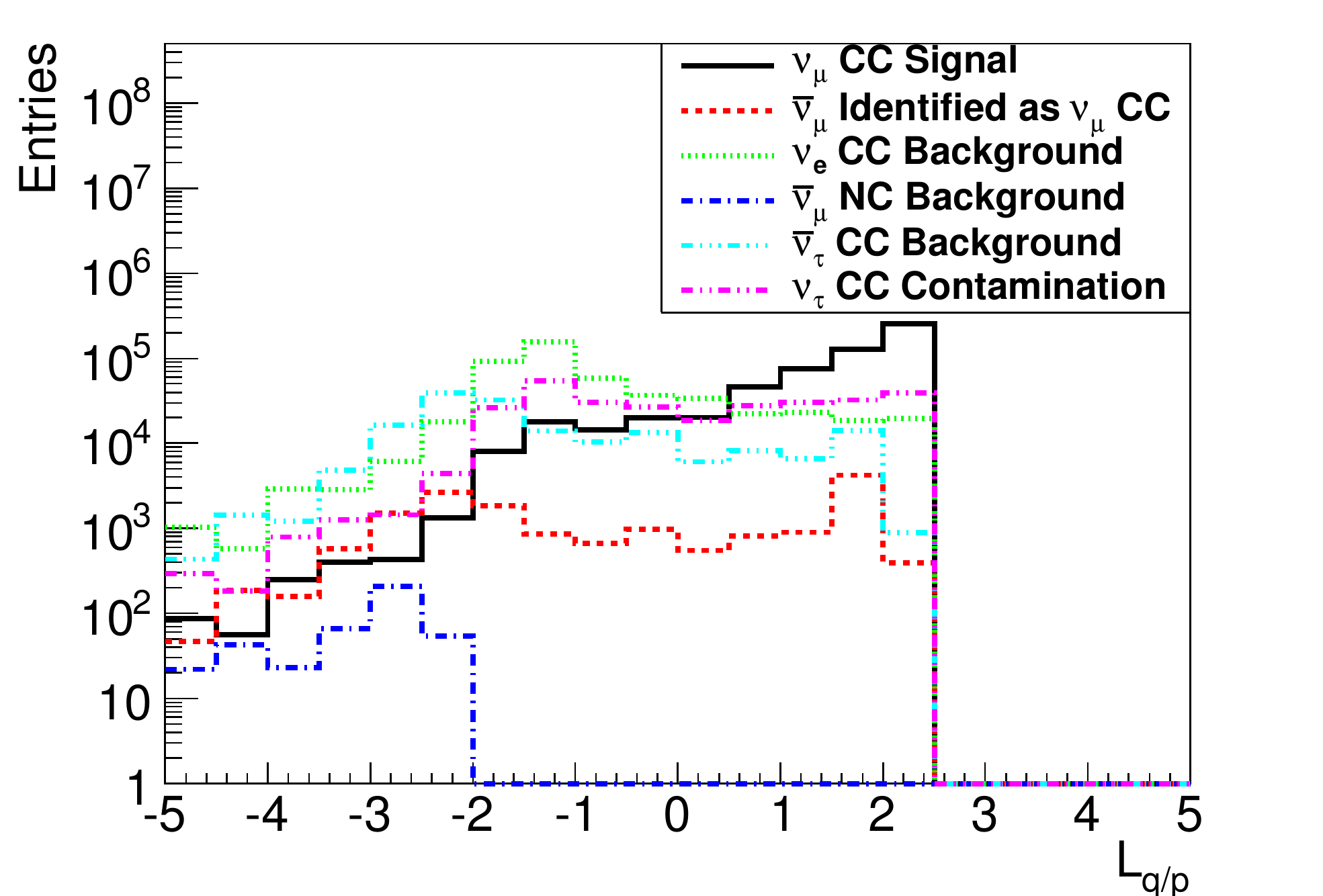}
}
\subfigure[Stored $\mu^-$ experiment]{
\includegraphics[width=0.475\textwidth]{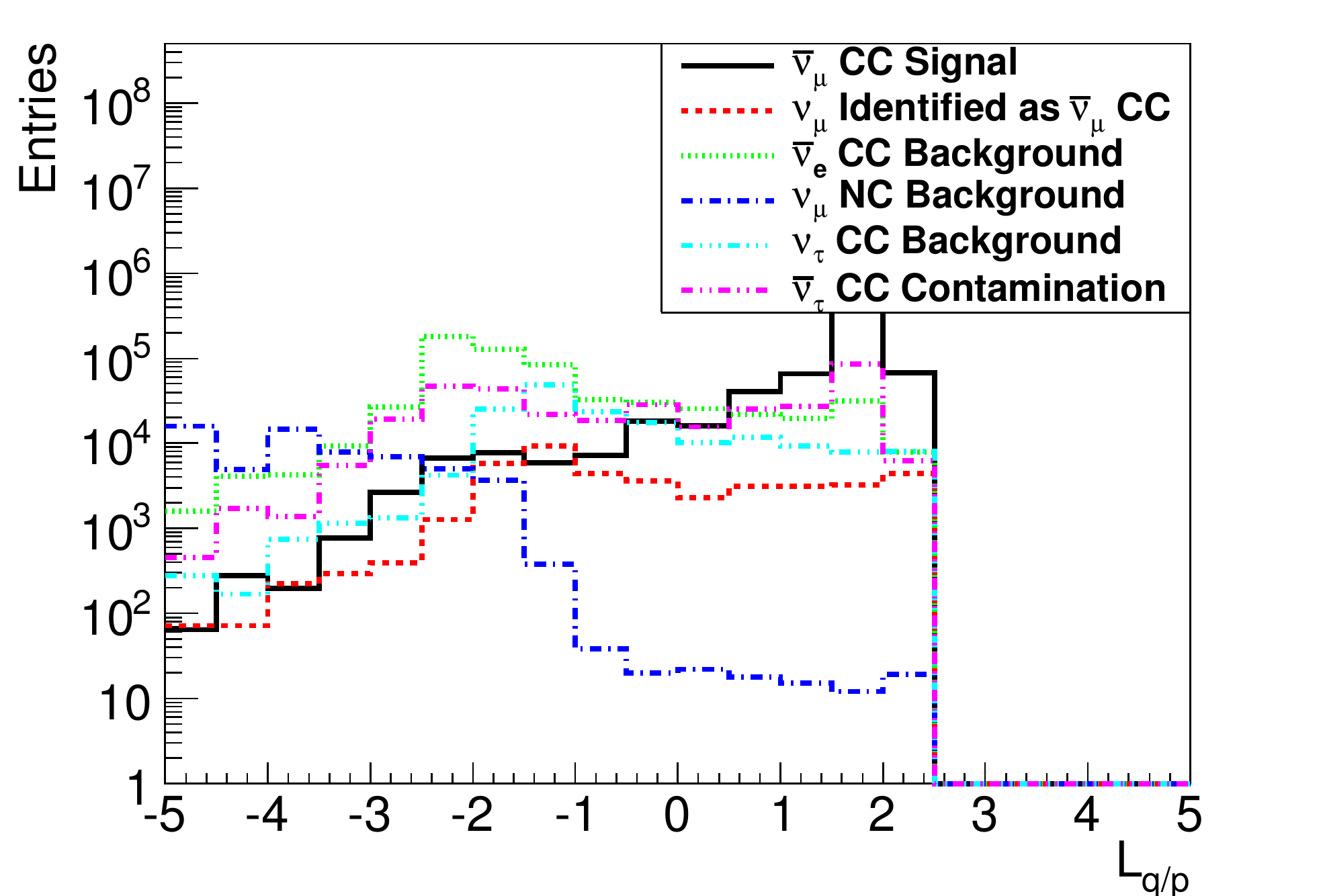}
}
\caption{Distribution of $\mathcal{L}_{q/p}$ for simulated neutrino species.}
\label{fig:Lqp}
\end{center}
\end{figure}

Two cuts deserve special attention as they provide most of the
discriminating power between the Golden channel oscillation signal and
background events. Both use a log-likelihood approach to select
between charge current and neutral current interactions. The
likelihood derived from the probability of the normalized uncertainty
in $q/p$ (the charge over the momentum from the fit to each track) for
charge current events $P(\sigma_{q/p}/(q/p) | CC)$ with respect to the
same for neutral current events $P(\sigma_{q/p}/(q/p) | NC)$ is
\begin{equation}
\mathcal{L}_{q/p} = \log\left(\frac{P(\sigma_{q/p}/(q/p) | CC)}{P(\sigma_{q/p}/(q/p) | NC)}\right),
\end{equation}
which provides good separation between signal and background when
$\mathcal{L}_{q/p} > -0.5$. This cut was chosen through consideration
of the distributions of $\mathcal{L}_{q/p}$ for the simulated neutrino
species as shown in Fig.~\ref{fig:Lqp}.

\begin{figure}[htbp]
\begin{center}
\subfigure[Stored $\mu^+$ experiment]{
\includegraphics[width=0.475\textwidth]{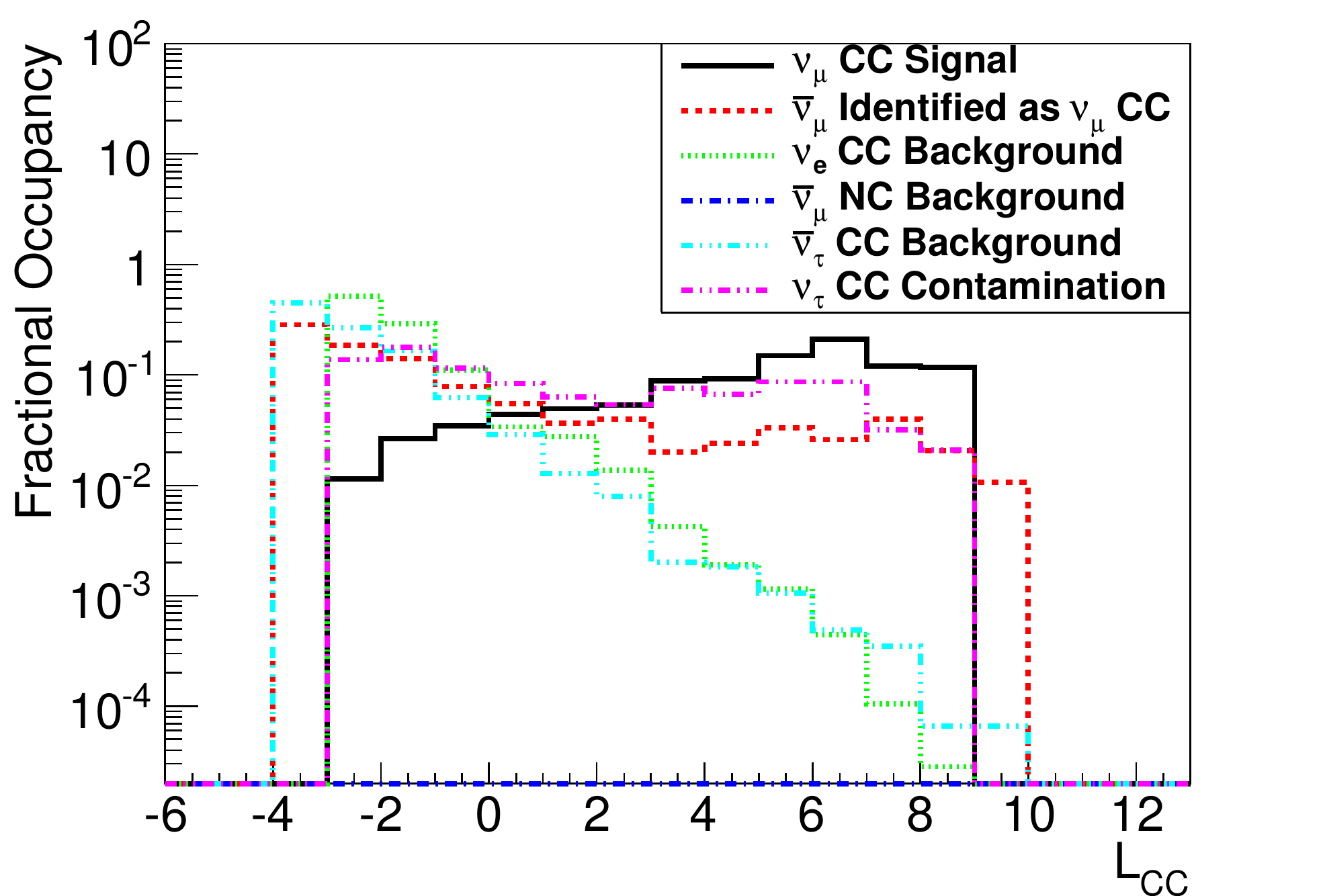}
}
\subfigure[Stored $\mu^-$ experiment]{
\includegraphics[width=0.475\textwidth]{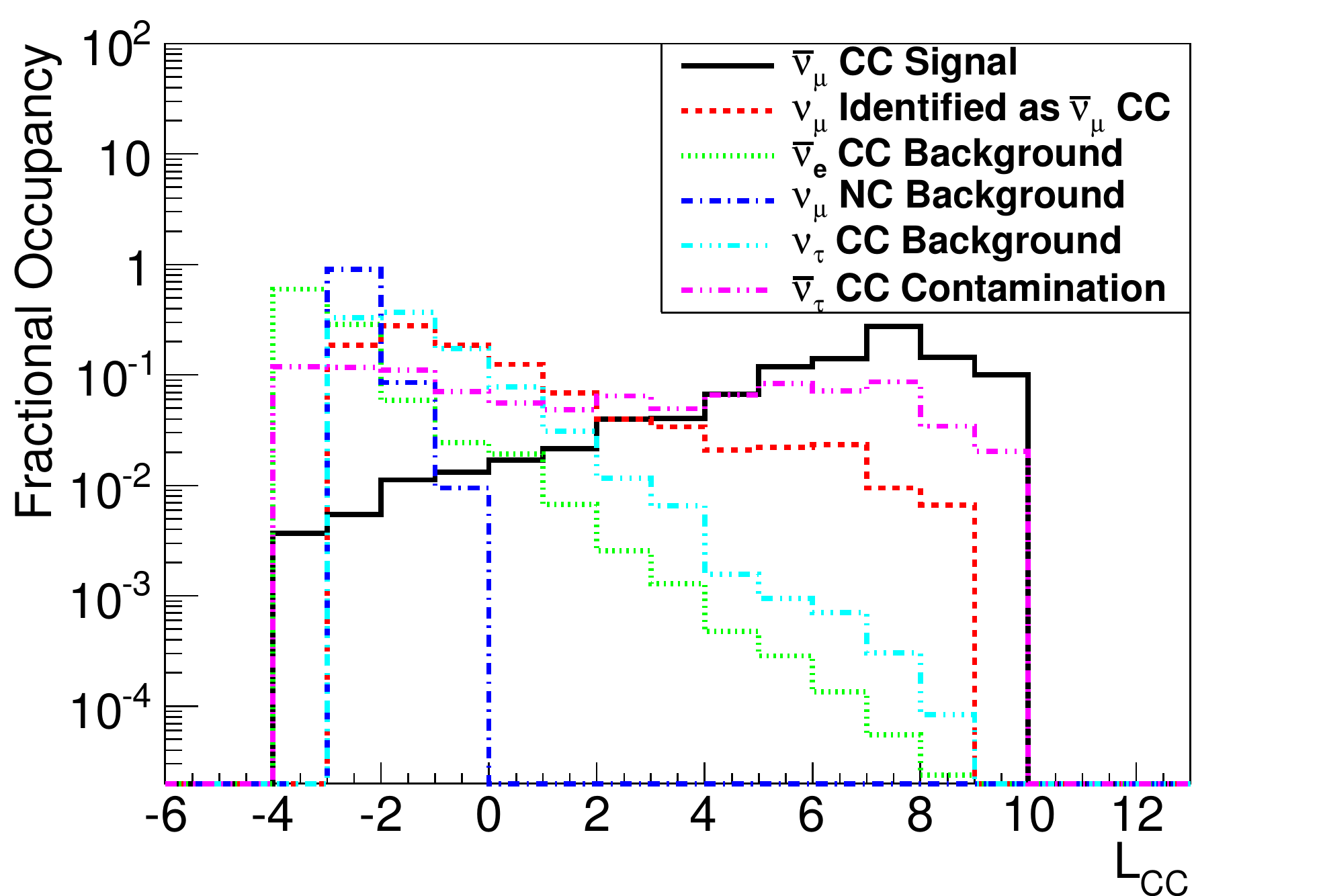}
}
\caption{Distribution of $\mathcal{L}_{CC}$ for simulated neutrino species.}
\label{fig:LCC1}
\end{center}
\end{figure}

A stronger charge current selection is defined by the number of hits
in the track. Muon tracks travel much further within the detector, so
they produce many more hits than electron or hadron showers, which are
known to range out quickly. To make this cut without bias a likelihood
ratio was defined as the probability of a track appearing with a given
number of hits assuming a charge current event $P(N_{hit} | CC)$ to
the same probability assuming a neutral current event $P(N_{hit} |
CC)$;
\begin{equation}
\mathcal{L}_{CC} =  \log\left(\frac{P(N_{hit} | CC)}{P(N_{hit}| NC)}\right).
\end{equation}
The best separation between signal and background occurs when events
with $\mathcal{L}_{CC} > 1.0$ are kept. The $\mathcal{L}_{CC}$
distributions for the simulated neutrino species are shown in
Fig.~\ref{fig:LCC1}.

\begin{figure}[htbp]
\begin{center}
\subfigure[Stored $\mu^+$ experiment]{
\includegraphics[width=0.475\textwidth]{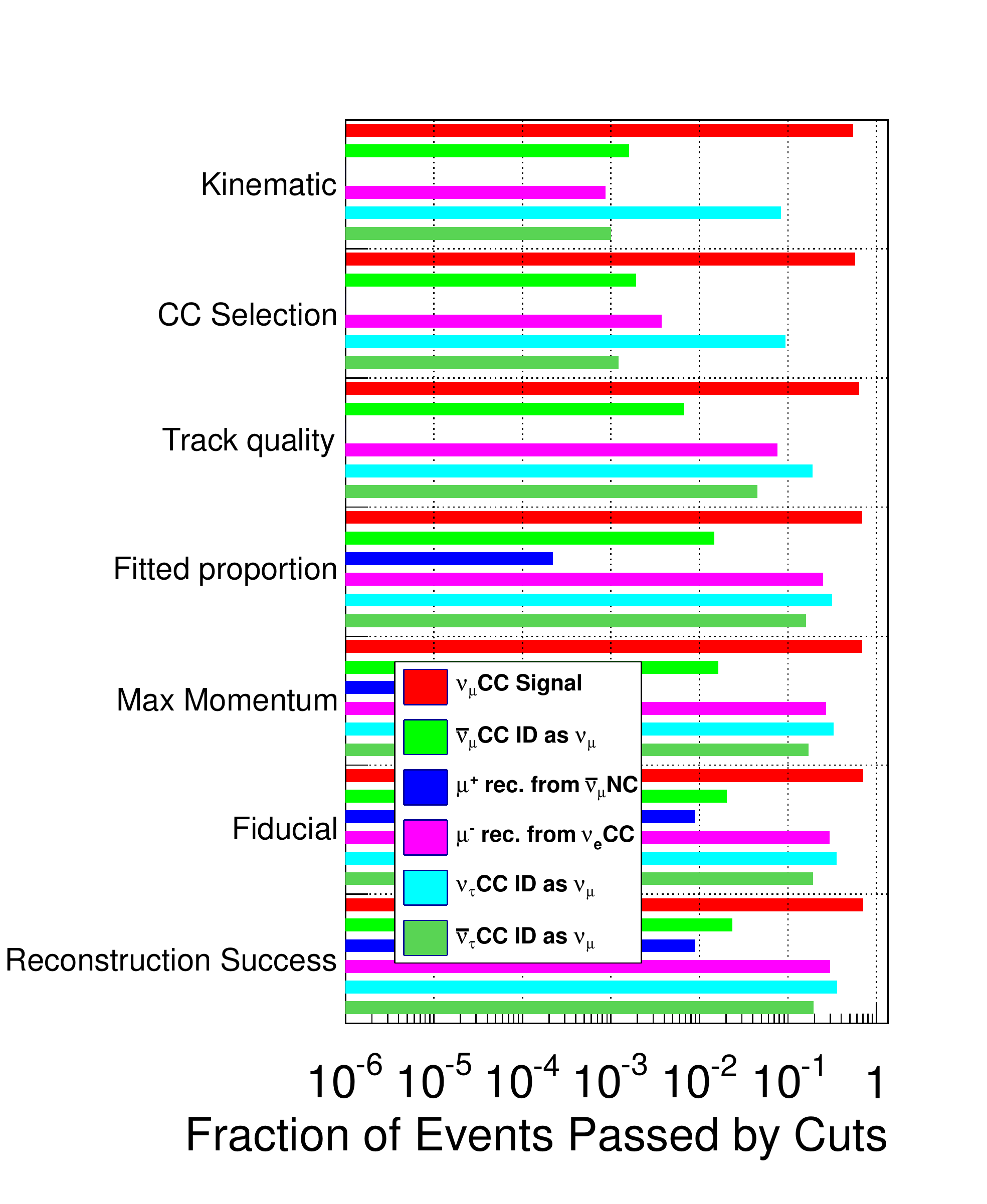}
}
\subfigure[Stored $\mu^-$ experiment]{
\includegraphics[width=0.475\textwidth]{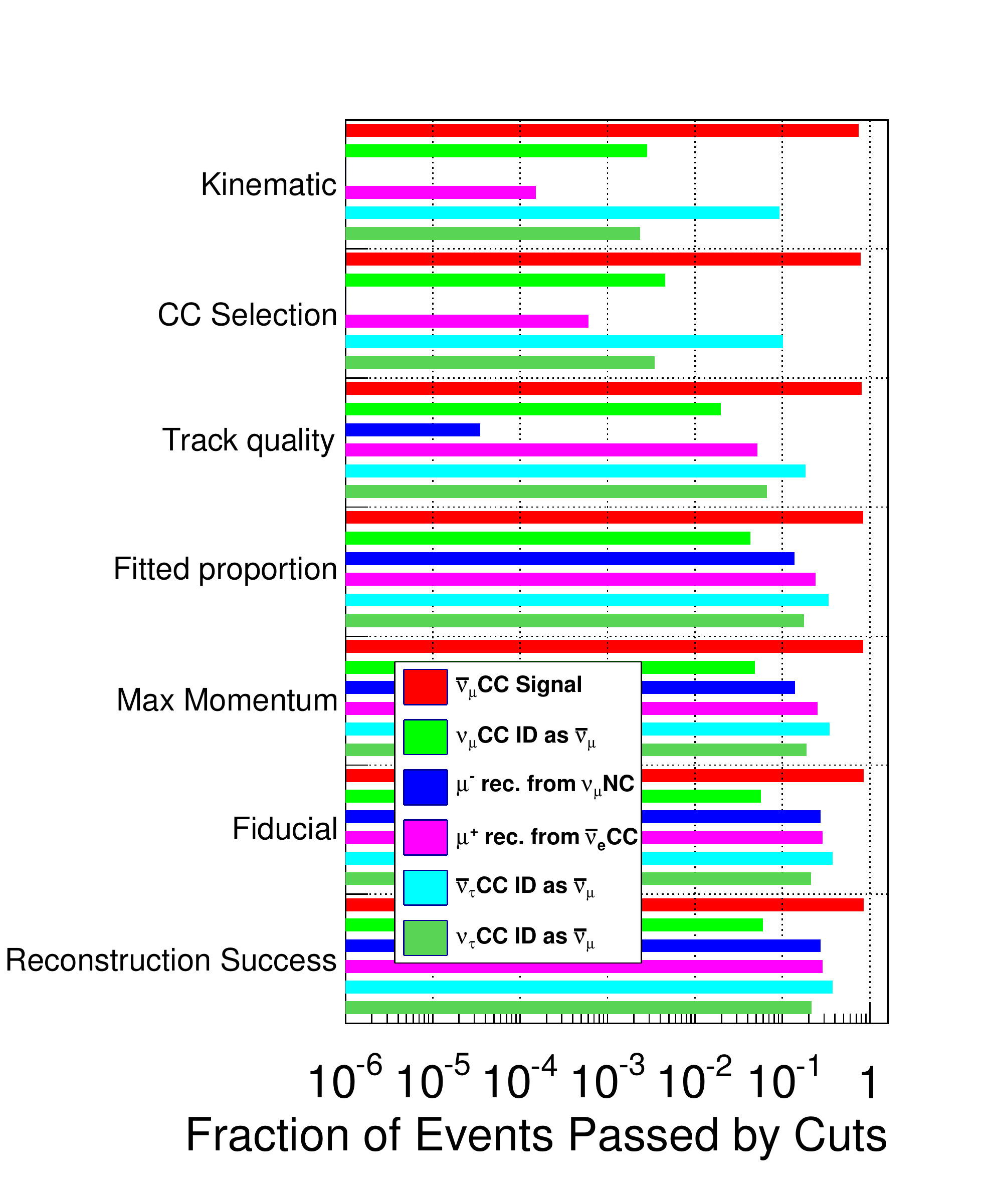}
}
\caption{Effect of the cuts on the detection of signal and background events assuming a magnetic field that focusses positively charged particles.}
\label{fig:cutseff_posfoc}
\end{center}
\end{figure}

\begin{figure}[htbp]
\begin{center}
\subfigure[Stored $\mu^+$ experiment]{
\includegraphics[width=0.475\textwidth]{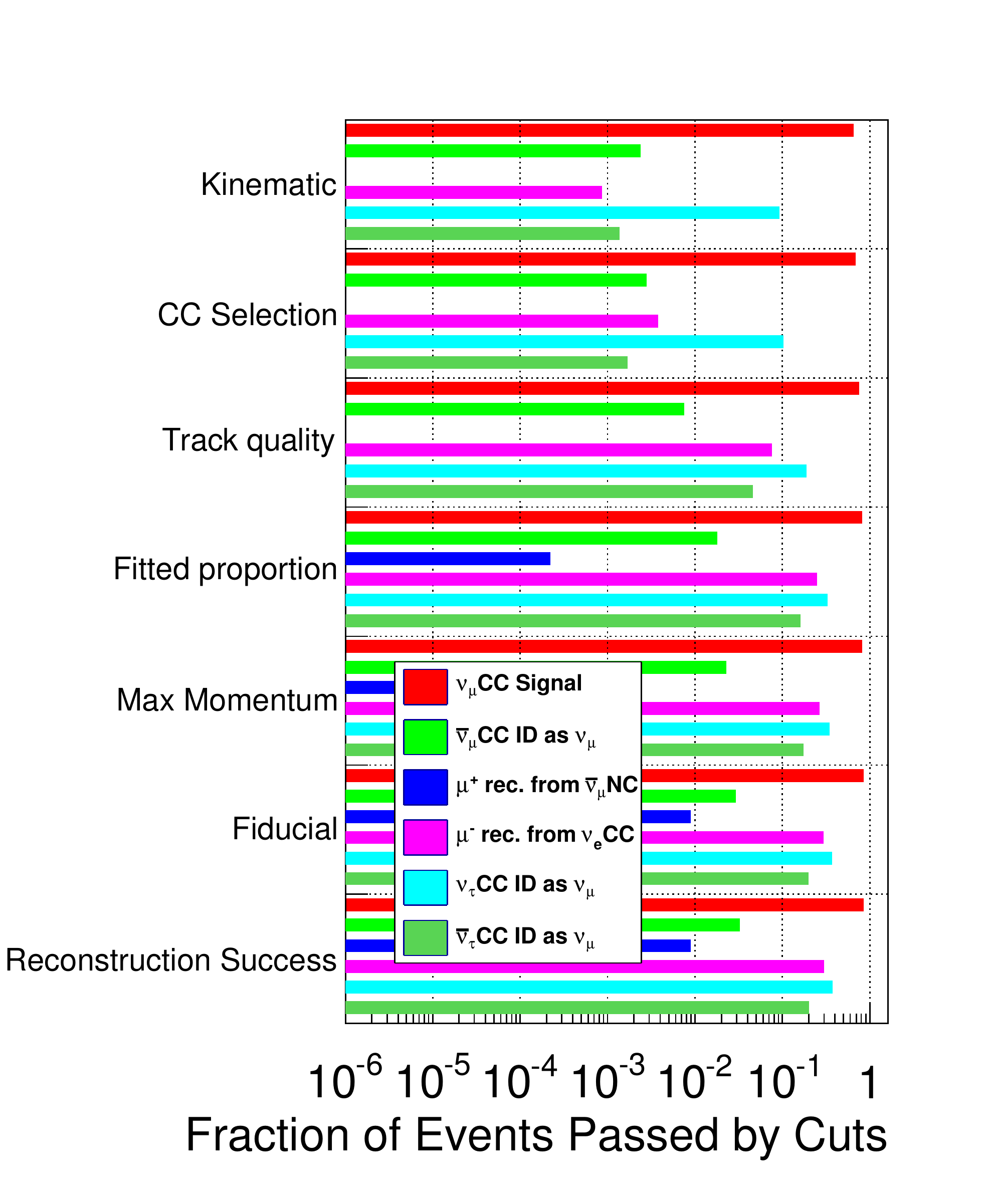}
}
\subfigure[Stored $\mu^-$ experiment]{
\includegraphics[width=0.475\textwidth]{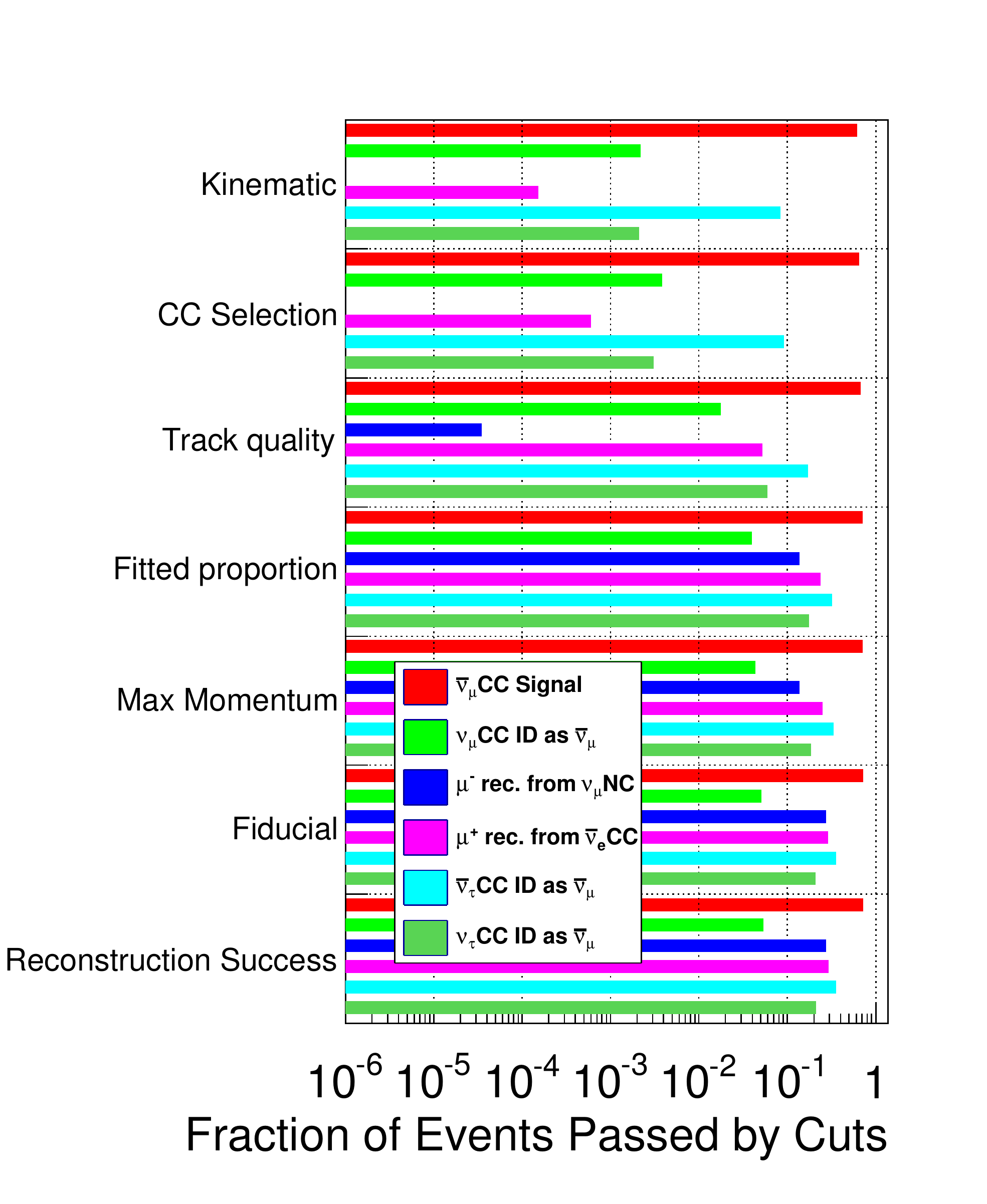}
}
\caption{Effect of the cuts on the detection of signal and background events assuming a magnetic field that focusses negatively charged particles.}
\label{fig:cutseff_negfoc}
\end{center}
\end{figure}

A further cut is applied on the kinematic variables of the event. By
placing a cut on the separation of the muon direction and the
direction of hadronization, some further separation between signal
events and $\nu_{e}$ CC events can be achieved. For a 10 GeV neutrino
factory a cut on the separation variable $Q_t = E_{\nu}\sin\theta >
0.15$~GeV, where $\theta$ is the angle between the muon candidate and
the hadronic-jet vector, was found to provide this.  The effect of the
cuts on the event samples is summarized in
Fig.~\ref{fig:cutseff_posfoc} and \ref{fig:cutseff_negfoc}.

A multi-variate analysis for the identification of $\nu_{\mu}(\bar{\nu}_{\mu})$ CC events is also under consideration. Based on the experience of the MINOS experiment \cite{Adamson:2011fa} it is believed that an analysis, such as a $k$-nearest neighbour approach, using multiple correlated variables can produce a better discrimination between signal and background events. A set of variables that includes the mean energy deposition along the muon track, the variation of the energy deposition, and the total number of track hits is under consideration for this purpose. This is work in progress.

\section{Detector Efficiencies and Response}\label{sec:Eff}

\begin{figure}[htbp]
\begin{center}
\subfigure[Signal detection efficiency in a $\mu^+$ focussing field]{
\includegraphics[width=0.475\textwidth]{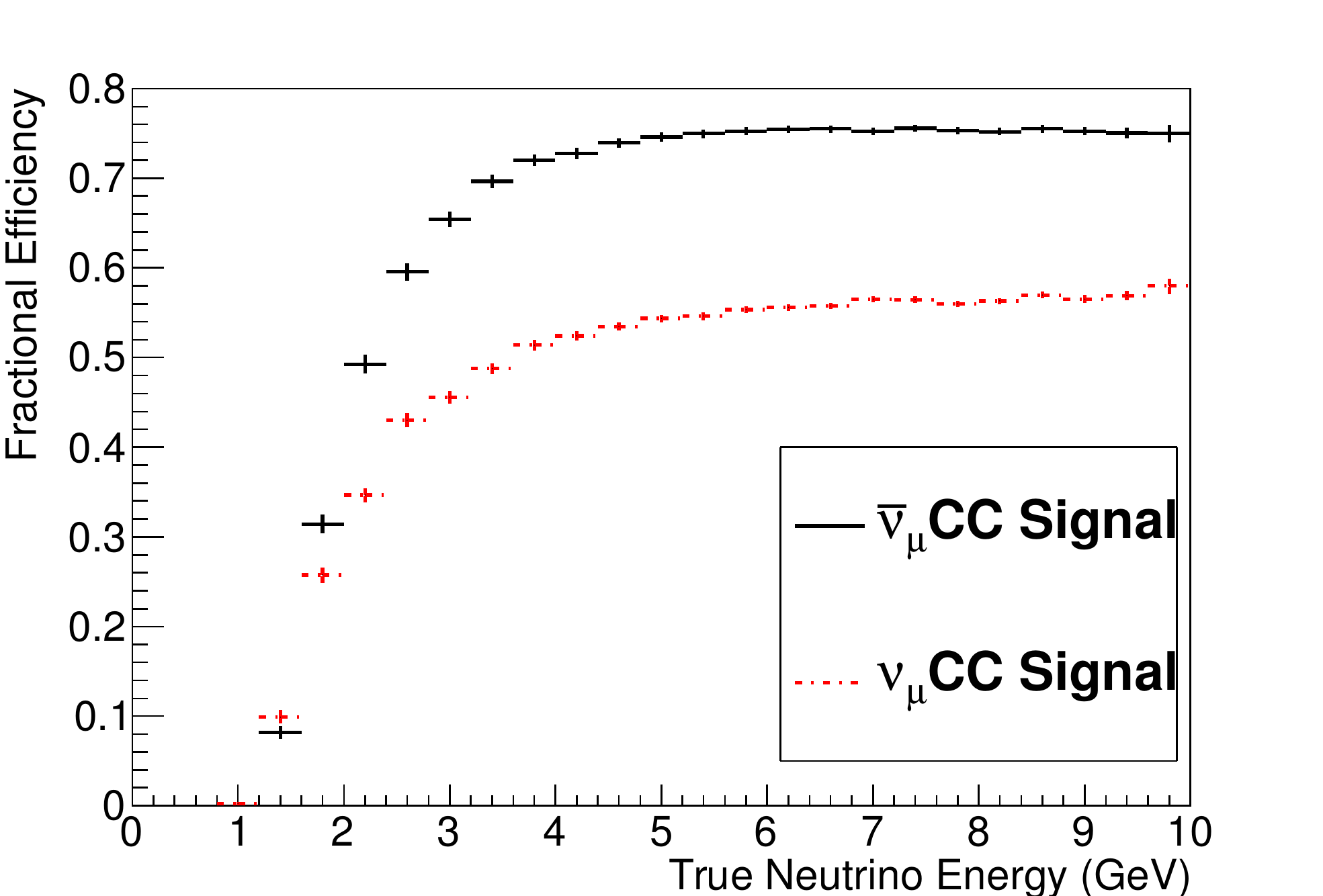}
\label{fig:mupluseff}
}
\subfigure[Signal detection efficiency in a $\mu^-$ focussing field]{
\includegraphics[width=0.475\textwidth]{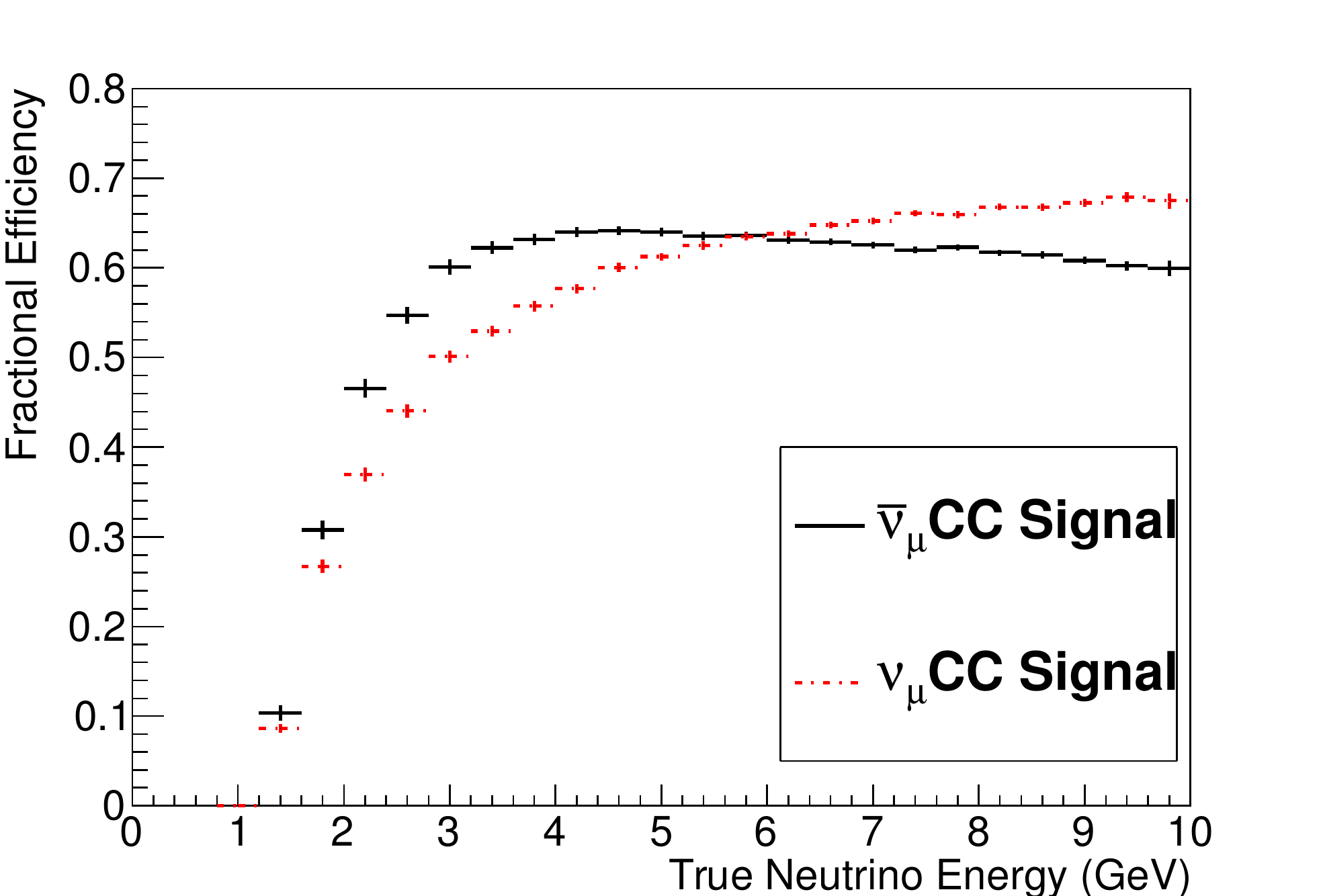}
\label{fig:muminuseff}
}
\caption{Efficiency for the detection of $\nu_{\mu}$ and $\bar{\nu}_{\mu}$ charge current detection assuming the  two different magnetic field configurations.}
\label{fig:eff}
\end{center}
\end{figure}

\begin{figure}[htbp]
\begin{center}
\subfigure[Background rates in a $\mu^+$ focussing field]{
\includegraphics[width=0.475\textwidth]{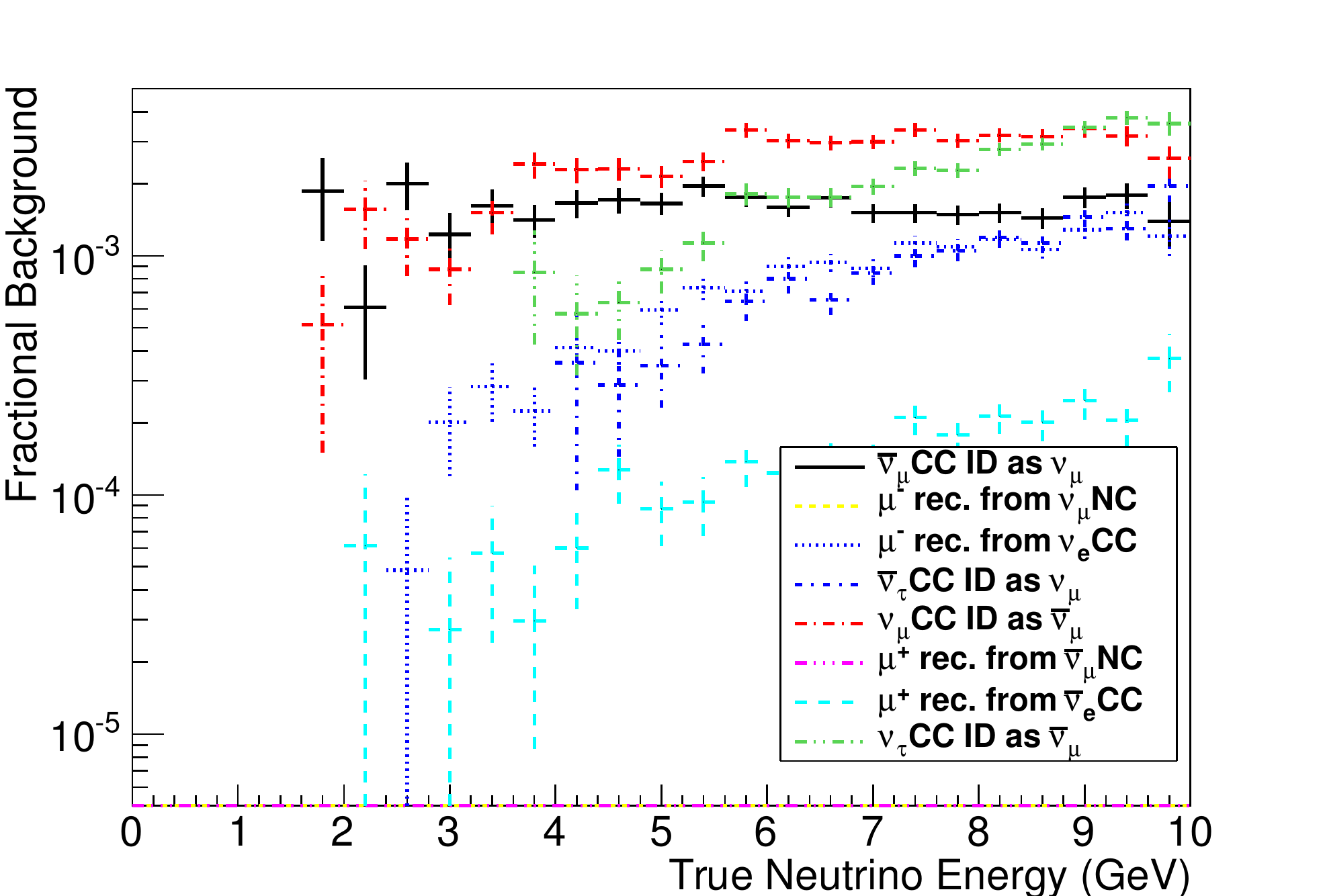}
\label{fig:muplusback}
}
\subfigure[Background rates in a $\mu^-$ focussing field]{
\includegraphics[width=0.475\textwidth]{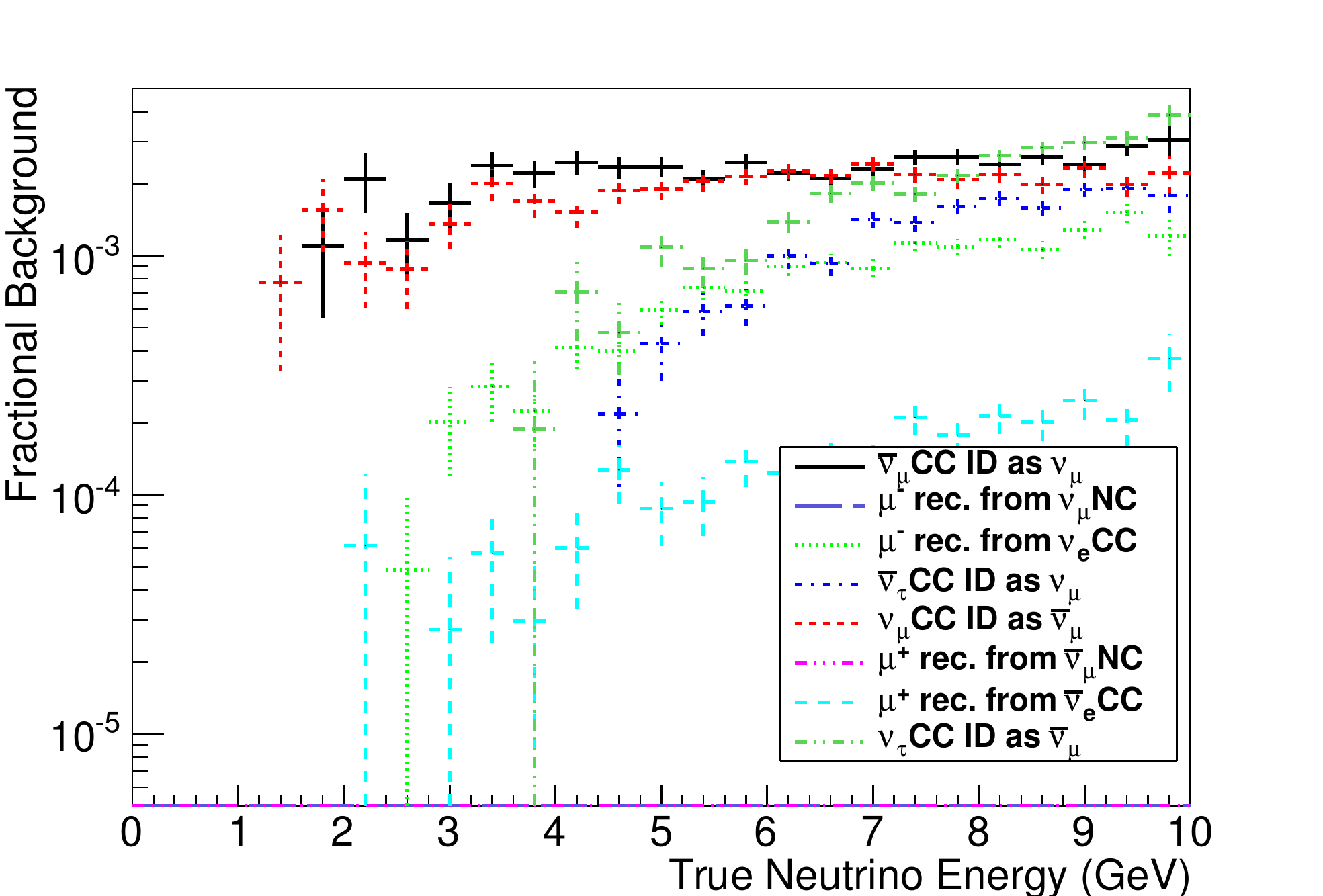}

\label{fig:muminusback}
}
\caption{Background rates for $\nu_{\mu}$ and $\bar{\nu}_{\mu}$ charge current detection assuming the  two different magnetic field configurations.}
\label{fig:back}
\end{center}
\end{figure}

The efficiencies and background suppression for MIND in the muon neutrino appearance channel is shown in Fig.~\ref{fig:eff} and Fig.~\ref{fig:back}. Four cases are considered in these figures depending on the binary state of the storage ring and the detector; a $\mu^{-}(\mu^{+})$ is contained in the storage ring resulting in a $\bar{\nu}_{\mu}(\nu_{\mu})$ signal, and the magnetic field of MIND focusses $\mu^{+}(\mu^{-})$. A neutrino factory stores both $\mu^{-}$ and $\mu^{+}$ in the ring so that pulses of neutrinos associated with decays of each species can be identified based on their correlated time structure. However the magnetic field direction must be chosen a priori based on an understanding of the detector response and resulting sensitivity to the CP violation. 
 
\section{Sensitivities}\label{sec:sense}

The analysis of the simulation is used to generate ``migration matrices" that relate the true neutrino energy to the reconstructed neutrino energy and contain all of the information regarding the reconstruction efficiency, energy response, and resolution.  These migration matrices ($M_{ij}$) are used to convert a set of neutrino counts  ($\nu(E_{j})$) calculated using a long baseline simulation into expected counts in a detector as a function of energy ($n(E_{i})$) i.e. $n(E_{i}) = M_{ij}\nu(E_{j})$. The Neutrino tool suite (NuTS), developed for the studies presented in~\cite{BurguetCastell:2001ez,BurguetCastell:2002qx,BurguetCastell:2005pa}, is  a framework that generates the event rates $\nu(E_{j})$ from the appropriate fluxes and is used to extract the neutrino oscillation probabilities for all channels. 

The pseudo-experimental data is extracted from a combination of the signal and background species,
\begin{equation}
n^{data}_i = M^{sig}_{ij}\nu^{sig}(E_{j}) + \sum_{k}M^{bkg,k}_{ij}\nu^{bkg,k}(E_j),
\end{equation}
and compared with an oscillation hypothesis using a $\chi^{2}$ statistic such as
\begin{eqnarray}\label{eq:chi2}
\chi^2 &=  2 \displaystyle \sum_{i = 0}^{L}\Bigg( & A x N_{+,i} (\theta_{13}, \delta_{CP}) - n^{data}_{+,i} + n^{data}_{+,i} \ln\left(\frac{n^{data}_{+,i}}{AxN_{+,i}(\theta_{13},\delta_{CP})}\right)\nonumber \\
& &+ A N_{-,i}(\theta_{13},\delta_{CP}) - n^{data}_{-,i} + n^{data}_{-,i}\ln\left(\frac{n^{data}_{-,i}}{AN_{-,i}(\theta_{13},\delta_{CP})}\right) \nonumber \\ & & +  \frac{(A - 1)^{2}}{\sigma_{A}^2} + \frac{(x - 1)^2}{\sigma_x} \Bigg).
\end{eqnarray}

In this equation $n^{data}_{q,i}$ is the simulated ``data" for the energy bin, $i$, assuming a muon signal with a sign $q$, while $N_{q,i}(\theta_{13},\delta_{CP})$ is the predicted content of the corresponding energy bin for the test values of $\theta_{13}$ and $\delta_{CP}$. This fit includes two systematic uncertainties that are assumed to be the leading terms; the error, $\sigma_{x}$, on the ratio $x$ of $\nu$ to $\bar{\nu}$ cross-sections and the error, $\sigma_{A}$, on the total counts in the detector $A$ due to fiducial errors or variation in the neutrino beam.  The uncertainty in the cross-section ratio is assumed to be measurable to the 1\% level at a neutrino factory, as the near detector sites will take concurrent measurements of both the neutrino and anti-neutrino species \cite{NearDetector:2013}. Similarly, measurements at the near detector combined with muon decay rate measurements from instrumentation in the muon decay ring should reduce the uncertainty $\sigma_{A}$ to below 1\% \cite{Bayes:2012ex}. Conservative upper limits for these errors of 3\% and 2.5\% respectively are also considered in this study, but the neutrino factory will allow much better control of these systematic uncertainties.   

\begin{figure}[htbp]
\begin{center}
\includegraphics[width=0.7\textwidth]{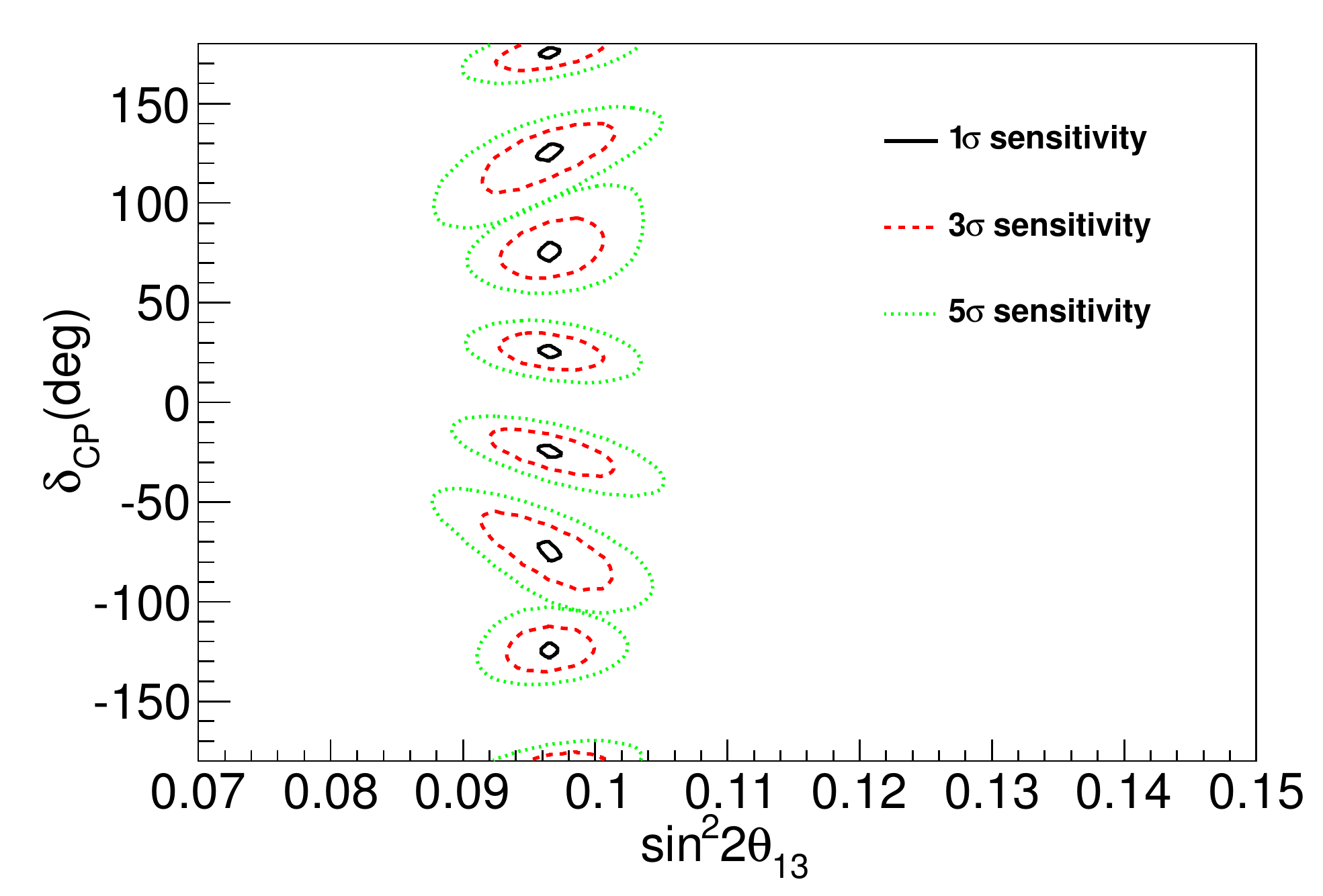}
\caption{Contours of constant $\chi^2$ for a variety of values of $\delta_{CP}$, calculated using Eq.~\ref{eq:chi2}. The simulations were generated with the octagonal geometry and a $\mu^{-}$ focussing field. Systematic errors of 1\% for both $\sigma_A$ and $\sigma_x$ are assumed in these fits.}
\label{fig:fit}
\end{center}
\end{figure}

\begin{figure}[htbp]
\begin{center}
\subfigure[Error assuming a $\mu^{-}$ focussing field.]{
	\includegraphics[width=0.475\textwidth]{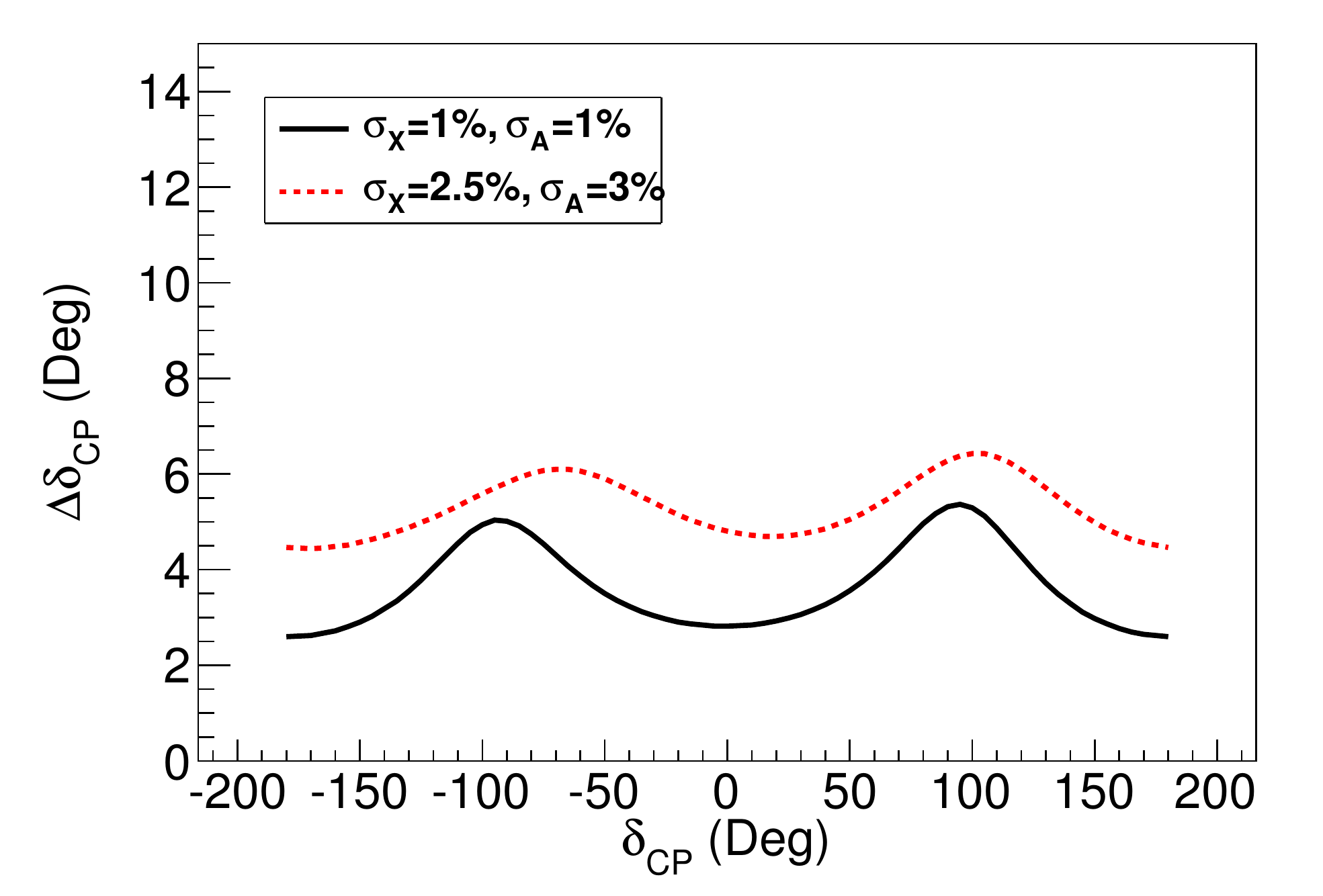}
}
\subfigure[Error assuming a $\mu^{+}$ focussing field.]{
	\includegraphics[width=0.475\textwidth]{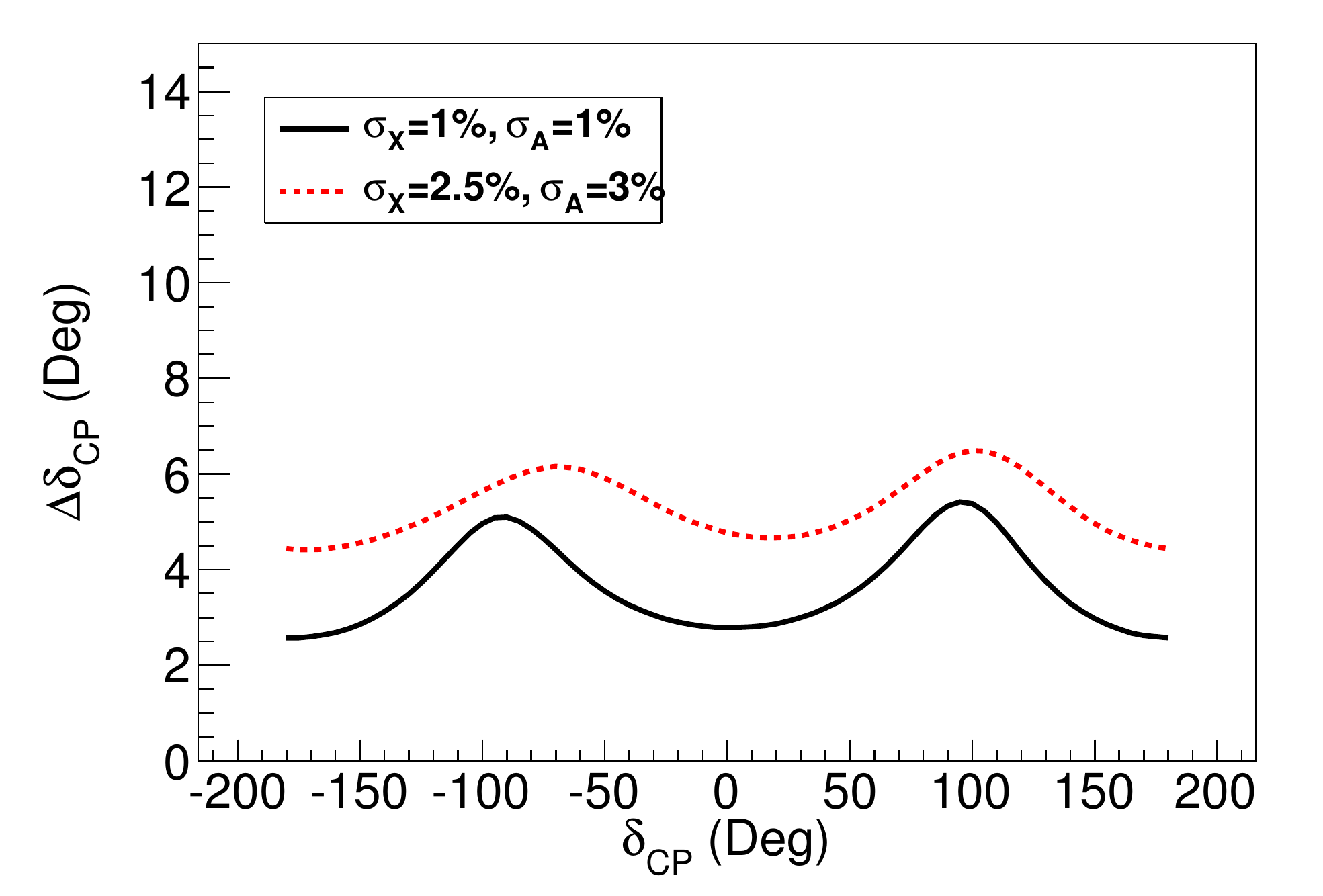}
}
\caption{Uncertainty on $\delta_{CP}$ as a function of $\delta_{CP}$ for the two possible detector field polarities. The black line assumes the 1\% systematic uncertainties that are believed will be prevalent at the neutrino factory, while the red line assumes uncertainties inflated by a factor of between 2.5 and 3. }
\label{fig:error}
\end{center}
\end{figure}

The $\chi^2$ contours defined by a series of fits with different values of CP violating phase are shown in Fig.~\ref{fig:fit}. The error of a measurement of the CP violating phase is determined by finding the width of the contour defined by $\chi^2 = 1$ at $\theta_{13}=9.0^{\circ}$. A neutrino factory offers  the best prospect to improve the precision of the measurement of $\theta_{13}$. The uncertainty curves for $\delta_{CP}$ derived from simulations using the $\mu^{+}$ and $\mu^{-}$ focusing fields are shown in Fig.~\ref{fig:error}. Uncertainty curves are identical for the two cases, suggesting that the variation in the momentum response resulting from the change in the detector field properties averages out when the species are added together for the $\chi^2$ calculation.

\begin{figure}[htbp]
\begin{center}
\subfigure[Sensitivity to CP violation, normal hierarchy]{
	\includegraphics[width=0.475\textwidth]{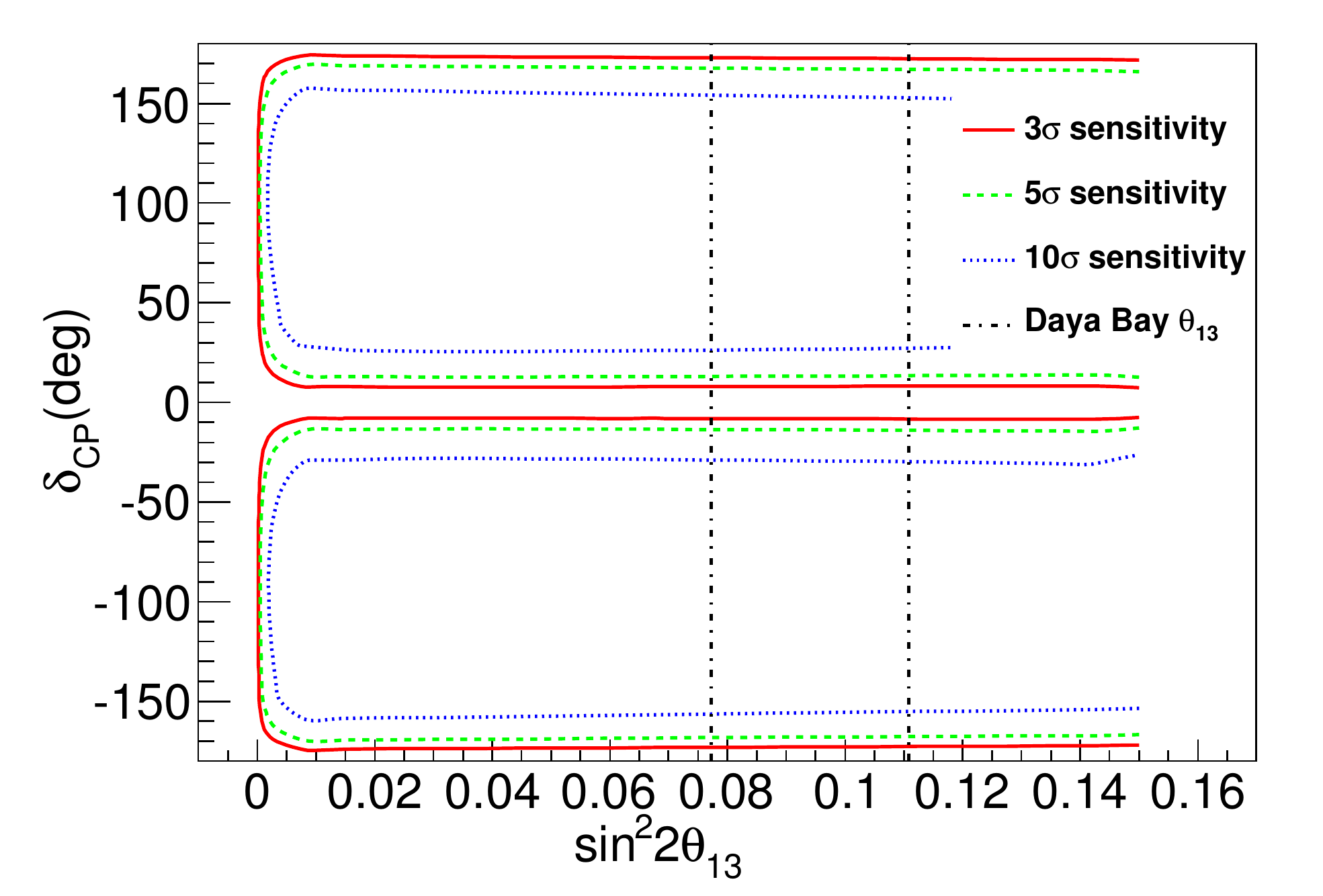}
}
\subfigure[Fractional coverage of 5$\sigma$ sensitivity, normal hierarchy]{
	\includegraphics[width=0.475\textwidth]{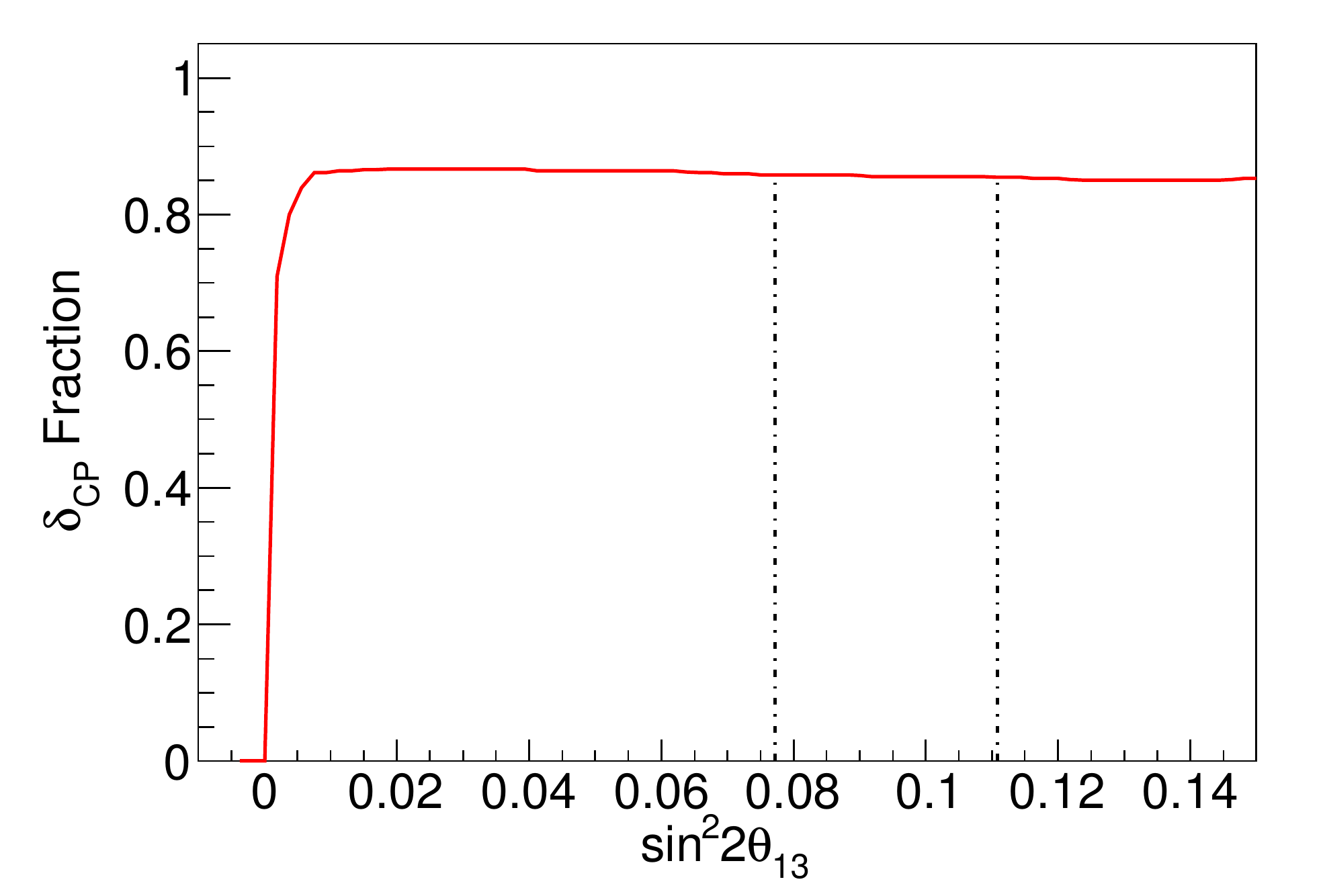}
}	
\subfigure[Sensitivity to CP violation,inverted hierarchy]{
	\includegraphics[width=0.475\textwidth]{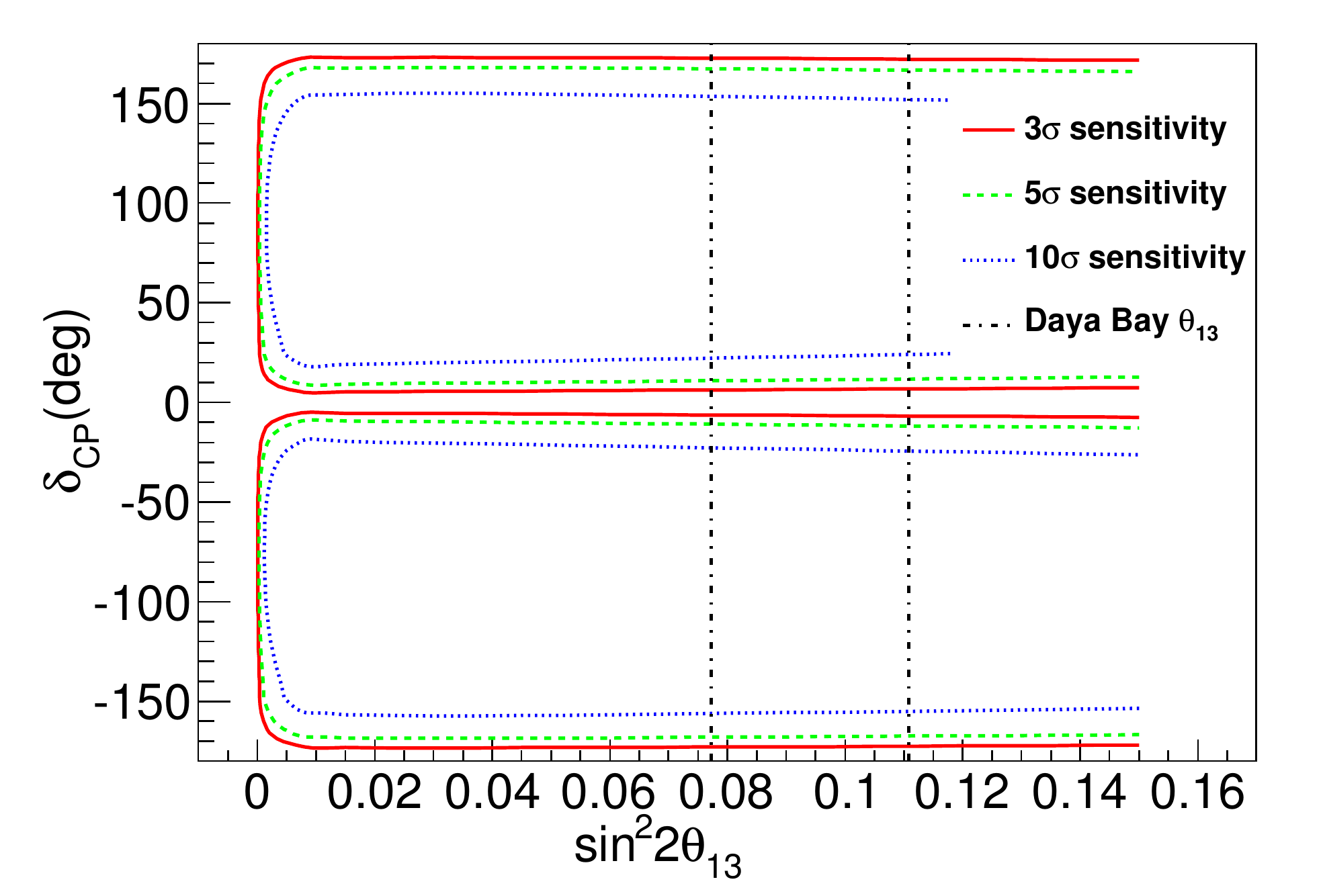}
}
\subfigure[Fractional coverage of 5$\sigma$ sensitivity, inverted hierarchy]{
	\includegraphics[width=0.475\textwidth]{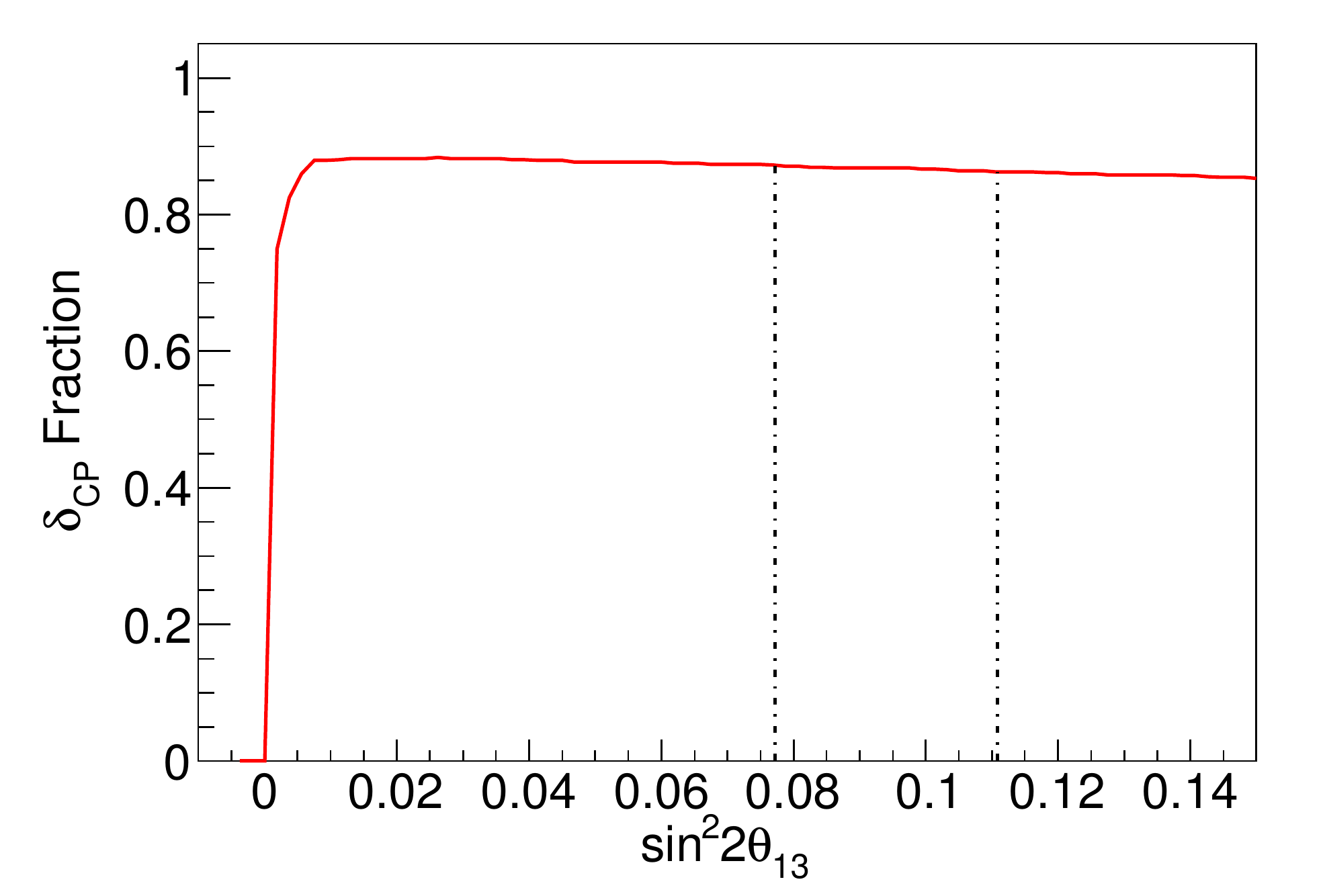}
}	

\caption{Sensitivity of a MIND at a neutrino factory to the discovery of CP violation assuming both a normal and inverted mass hierarchy as noted. The range of $\sin^2 2\theta_{13}$ measured by Daya Bay is shown with the black dotted line superimposed on the coverage contours and the fractional 5$\sigma$ coverage.}
\label{fig:sens}
\end{center}
\end{figure}

The sensitivity of the neutrino factory to CP violation can be determined by searching for sets of oscillation parameters that satisfy the inequality
\begin{equation}\label{eq:sens}
\max(\chi^{2}(\delta_{CP}=-180^{\circ}),\chi^{2}(\delta_{CP}=0^{\circ}),\chi^{2}(\delta_{13}^{0}=180^{\circ})) - \chi^{2}_{min} \geq n^2
\end{equation}
where $n$ is the desired significance level for the calculation. The curves showing the sensitivity to CP violation and the corresponding fractional 5$\sigma$ coverage are shown in Fig.~\ref{fig:sens}. A neutrino factory  can measure 85\% of the possible values of $\delta_{CP}$ within the measured range of $\theta_{13}$ with a 5$\sigma$ significance. No change in these figures results from changing the polarity of the detector field.

A similar inequality to that shown in Eq.~\ref{eq:sens} can be defined for the sensitivity to the mass hierarchy. The 5$\sigma$ mass hierarchy  discovery potential is achieved for all values of $\theta_{13}$ in which $\sin^{2}2\theta_{13}>10^{-4}$, therefore a neutrino factory will be sensitive to the mass hierarchy for the currently measured value of $\theta_{13}$.  

\section{Conclusions}

A detailed simulation of a  magnetized iron detector with a toroidal field has been produced for neutrino factory studies. This simulation shows that the neutrino factory is capable of discovering CP violation for 85\% of the values of the CP violating phase. This result is independent of the mass hierarchy. Given the recent measurements of $\theta_{13}$ by Daya Bay and others, the precision of the measurement is determined to be between $2.5^{\circ}$ and $5.5^{\circ}$ depending on the value of the CP violating phase and assuming leading systematic errors of 1\%. Should the sum of the systematics increase to 3.5\%, then the largest uncertainty on the measurement of $\delta_{CP}$ is 7$^{\circ}$. These results assume a measurement based on 5$\times 10^{21}$ muons of both species collected over 10 years. 

Further work is in progress to refine these results. Improvement in the reconstruction of multiple tracks for the purpose of identifying hadron showers is in progress and will be implemented soon. Likewise, a multi-variate analysis of the reconstructed simulation is under development. Other studies of the behaviour of MIND will become priorities after the completion of these developments including systematic studies and investigation of the impact of cosmic rays. These studies will come to a conclusion prior to the neutrino factory reference design report due at the end of 2013.

\begin{acknowledgments}
The authors acknowledge the support of the European Community under
the European Commission Framework Programme 7 Design Study: EUROnu,
Project Number 212372. 
The work was supported by the Science and Technology Facilities
Council (UK), the Spanish Ministry of Education and Science and the Department of Energy (USA).  
\end{acknowledgments}
\bibliography{MINDStudies}

\begin{thebibliography}{29}%
\makeatletter
\providecommand \@ifxundefined [1]{%
 \@ifx{#1\undefined}
}%
\providecommand \@ifnum [1]{%
 \ifnum #1\expandafter \@firstoftwo
 \else \expandafter \@secondoftwo
 \fi
}%
\providecommand \@ifx [1]{%
 \ifx #1\expandafter \@firstoftwo
 \else \expandafter \@secondoftwo
 \fi
}%
\providecommand \natexlab [1]{#1}%
\providecommand \enquote  [1]{``#1''}%
\providecommand \bibnamefont  [1]{#1}%
\providecommand \bibfnamefont [1]{#1}%
\providecommand \citenamefont [1]{#1}%
\providecommand \href@noop [0]{\@secondoftwo}%
\providecommand \href [0]{\begingroup \@sanitize@url \@href}%
\providecommand \@href[1]{\@@startlink{#1}\@@href}%
\providecommand \@@href[1]{\endgroup#1\@@endlink}%
\providecommand \@sanitize@url [0]{\catcode `\\12\catcode `\$12\catcode
  `\&12\catcode `\#12\catcode `\^12\catcode `\_12\catcode `\%12\relax}%
\providecommand \@@startlink[1]{}%
\providecommand \@@endlink[0]{}%
\providecommand \url  [0]{\begingroup\@sanitize@url \@url }%
\providecommand \@url [1]{\endgroup\@href {#1}{\urlprefix }}%
\providecommand \urlprefix  [0]{URL }%
\providecommand \Eprint [0]{\href }%
\providecommand \doibase [0]{http://dx.doi.org/}%
\providecommand \selectlanguage [0]{\@gobble}%
\providecommand \bibinfo  [0]{\@secondoftwo}%
\providecommand \bibfield  [0]{\@secondoftwo}%
\providecommand \translation [1]{[#1]}%
\providecommand \BibitemOpen [0]{}%
\providecommand \bibitemStop [0]{}%
\providecommand \bibitemNoStop [0]{.\EOS\space}%
\providecommand \EOS [0]{\spacefactor3000\relax}%
\providecommand \BibitemShut  [1]{\csname bibitem#1\endcsname}%
\let\auto@bib@innerbib\@empty
\bibitem [{\citenamefont {Geer}(1998)}]{Geer:1997iz}%
  \BibitemOpen
  \bibfield  {author} {\bibinfo {author} {\bibfnamefont {S.}~\bibnamefont
  {Geer}},\ }\href@noop {} {\bibfield  {journal} {\bibinfo  {journal} {Phys.
  Rev.}\ }\textbf {\bibinfo {volume} {D57}},\ \bibinfo {pages} {6989} (\bibinfo
  {year} {1998})},\ \Eprint {http://arxiv.org/abs/hep-ph/9712290}
  {hep-ph/9712290} \BibitemShut {NoStop}%
\bibitem [{\citenamefont {De~R\'ujula}\ \emph {et~al.}(1999)\citenamefont
  {De~R\'ujula}, \citenamefont {Gavela},\ and\ \citenamefont
  {Hern\'andez}}]{DeRujula:1998hd}%
  \BibitemOpen
  \bibfield  {author} {\bibinfo {author} {\bibfnamefont {A.}~\bibnamefont
  {De~R\'ujula}}, \bibinfo {author} {\bibfnamefont {M.~B.}\ \bibnamefont
  {Gavela}}, \ and\ \bibinfo {author} {\bibfnamefont {P.}~\bibnamefont
  {Hern\'andez}},\ }\href {\doibase 10.1016/S0550-3213(99)00070-X} {\bibfield
  {journal} {\bibinfo  {journal} {Nucl. Phys.}\ }\textbf {\bibinfo {volume}
  {B547}},\ \bibinfo {pages} {21} (\bibinfo {year} {1999})},\ \Eprint
  {http://arxiv.org/abs/hep-ph/9811390} {arXiv:hep-ph/9811390} \BibitemShut
  {NoStop}%
\bibitem [{\citenamefont {Cervera}\ \emph {et~al.}(2000)\citenamefont {Cervera}
  \emph {et~al.}}]{Cervera:2000kp}%
  \BibitemOpen
  \bibfield  {author} {\bibinfo {author} {\bibfnamefont {A.}~\bibnamefont
  {Cervera}} \emph {et~al.},\ }\href {\doibase 10.1016/S0550-3213(00)00221-2}
  {\bibfield  {journal} {\bibinfo  {journal} {Nucl. Phys.}\ }\textbf {\bibinfo
  {volume} {B579}},\ \bibinfo {pages} {17} (\bibinfo {year} {2000})},\ \Eprint
  {http://arxiv.org/abs/hep-ph/0002108} {arXiv:hep-ph/0002108} \BibitemShut
  {NoStop}%
\bibitem [{IDS()}]{IDS-NF}%
  \BibitemOpen
  \href@noop {} {\enquote {\bibinfo {title} {{The International Design Study
  for the Neutrino Factory}},}\ }\bibinfo {note} {{URL:
  https://www.ids-nf.org/wiki/FrontPage}}\BibitemShut {NoStop}%
\bibitem [{EUR()}]{EUROnu}%
  \BibitemOpen
  \href@noop {} {\enquote {\bibinfo {title} {{EUROnu: A High Intensity Neutrino
  Oscillation Facility in Europe}},}\ }\bibinfo {note} {{URL:
  http://www.euronu.org/}}\BibitemShut {NoStop}%
\bibitem [{\citenamefont {Choubey}\ \emph {et~al.}(2011)\citenamefont {Choubey}
  \emph {et~al.}}]{Choubey:2011zz}%
  \BibitemOpen
  \bibfield  {author} {\bibinfo {author} {\bibfnamefont {S.}~\bibnamefont
  {Choubey}} \emph {et~al.} (\bibinfo {collaboration} {IDS-NF Collaboration}),\
  }\href@noop {} {\  (\bibinfo {year} {2011})},\ \Eprint
  {http://arxiv.org/abs/1112.2853} {arXiv:1112.2853 [hep-ex]} \BibitemShut
  {NoStop}%
\bibitem [{\citenamefont {Michael}\ \emph {et~al.}(2008)\citenamefont {Michael}
  \emph {et~al.}}]{Michael:2008bc}%
  \BibitemOpen
  \bibfield  {author} {\bibinfo {author} {\bibfnamefont {D.}~\bibnamefont
  {Michael}} \emph {et~al.} (\bibinfo {collaboration} {MINOS Collaboration}),\
  }\href {\doibase 10.1016/j.nima.2008.08.003} {\bibfield  {journal} {\bibinfo
  {journal} {Nucl.Instrum.Meth.}\ }\textbf {\bibinfo {volume} {A596}},\
  \bibinfo {pages} {190} (\bibinfo {year} {2008})},\ \Eprint
  {http://arxiv.org/abs/0805.3170} {arXiv:0805.3170 [physics.ins-det]}
  \BibitemShut {NoStop}%
\bibitem [{\citenamefont {Abe}\ \emph {et~al.}(2011)\citenamefont {Abe} \emph
  {et~al.}}]{Abe:2011sj}%
  \BibitemOpen
  \bibfield  {author} {\bibinfo {author} {\bibfnamefont {K.}~\bibnamefont
  {Abe}} \emph {et~al.} (\bibinfo {collaboration} {T2K Collaboration}),\ }\href
  {\doibase 10.1103/PhysRevLett.107.041801} {\bibfield  {journal} {\bibinfo
  {journal} {Phys.Rev.Lett.}\ }\textbf {\bibinfo {volume} {107}},\ \bibinfo
  {pages} {041801} (\bibinfo {year} {2011})},\ \Eprint
  {http://arxiv.org/abs/1106.2822} {arXiv:1106.2822 [hep-ex]} \BibitemShut
  {NoStop}%
\bibitem [{\citenamefont {An}\ \emph {et~al.}(2012)\citenamefont {An} \emph
  {et~al.}}]{An:2012eh}%
  \BibitemOpen
  \bibfield  {author} {\bibinfo {author} {\bibfnamefont {F.}~\bibnamefont {An}}
  \emph {et~al.} (\bibinfo {collaboration} {DAYA-BAY Collaboration}),\ }\href
  {\doibase 10.1103/PhysRevLett.108.171803} {\bibfield  {journal} {\bibinfo
  {journal} {Phys.Rev.Lett.}\ }\textbf {\bibinfo {volume} {108}},\ \bibinfo
  {pages} {171803} (\bibinfo {year} {2012})},\ \Eprint
  {http://arxiv.org/abs/1203.1669} {arXiv:1203.1669 [hep-ex]} \BibitemShut
  {NoStop}%
\bibitem [{\citenamefont {Ahn}\ \emph {et~al.}(2012)\citenamefont {Ahn} \emph
  {et~al.}}]{Ahn:2012nd}%
  \BibitemOpen
  \bibfield  {author} {\bibinfo {author} {\bibfnamefont {J.}~\bibnamefont
  {Ahn}} \emph {et~al.} (\bibinfo {collaboration} {RENO collaboration}),\
  }\href {\doibase 10.1103/PhysRevLett.108.191802} {\bibfield  {journal}
  {\bibinfo  {journal} {Phys.Rev.Lett.}\ }\textbf {\bibinfo {volume} {108}},\
  \bibinfo {pages} {191802} (\bibinfo {year} {2012})},\ \Eprint
  {http://arxiv.org/abs/1204.0626} {arXiv:1204.0626 [hep-ex]} \BibitemShut
  {NoStop}%
\bibitem [{\citenamefont {Abe}\ \emph {et~al.}(2012)\citenamefont {Abe} \emph
  {et~al.}}]{Abe:2011fz}%
  \BibitemOpen
  \bibfield  {author} {\bibinfo {author} {\bibfnamefont {Y.}~\bibnamefont
  {Abe}} \emph {et~al.} (\bibinfo {collaboration} {DOUBLE-CHOOZ
  Collaboration}),\ }\href {\doibase 10.1103/PhysRevLett.108.131801} {\bibfield
   {journal} {\bibinfo  {journal} {Phys.Rev.Lett.}\ }\textbf {\bibinfo {volume}
  {108}},\ \bibinfo {pages} {131801} (\bibinfo {year} {2012})},\ \Eprint
  {http://arxiv.org/abs/1112.6353} {arXiv:1112.6353 [hep-ex]} \BibitemShut
  {NoStop}%
\bibitem [{\citenamefont {Adamson}\ \emph
  {et~al.}(2011{\natexlab{a}})\citenamefont {Adamson} \emph
  {et~al.}}]{Adamson:2011qu}%
  \BibitemOpen
  \bibfield  {author} {\bibinfo {author} {\bibfnamefont {P.}~\bibnamefont
  {Adamson}} \emph {et~al.} (\bibinfo {collaboration} {MINOS Collaboration}),\
  }\href {\doibase 10.1103/PhysRevLett.107.181802} {\bibfield  {journal}
  {\bibinfo  {journal} {Phys.Rev.Lett.}\ }\textbf {\bibinfo {volume} {107}},\
  \bibinfo {pages} {181802} (\bibinfo {year} {2011}{\natexlab{a}})},\ \Eprint
  {http://arxiv.org/abs/1108.0015} {arXiv:1108.0015 [hep-ex]} \BibitemShut
  {NoStop}%
\bibitem [{\citenamefont {Bayes}\ \emph {et~al.}(2012)\citenamefont {Bayes},
  \citenamefont {Laing}, \citenamefont {Soler}, \citenamefont
  {Cervera~Villanueva}, \citenamefont {G\'omez~Cadenas}, \citenamefont
  {Hern\'andez}, \citenamefont {Mart\'in-Albo},\ and\ \citenamefont
  {Burguet-Castell}}]{Bayes:2012ex}%
  \BibitemOpen
  \bibfield  {author} {\bibinfo {author} {\bibfnamefont {R.}~\bibnamefont
  {Bayes}}, \bibinfo {author} {\bibfnamefont {A.}~\bibnamefont {Laing}},
  \bibinfo {author} {\bibfnamefont {F.~J.~P.}\ \bibnamefont {Soler}}, \bibinfo
  {author} {\bibfnamefont {A.}~\bibnamefont {Cervera~Villanueva}}, \bibinfo
  {author} {\bibfnamefont {J.~J.}\ \bibnamefont {G\'omez~Cadenas}}, \bibinfo
  {author} {\bibfnamefont {P.}~\bibnamefont {Hern\'andez}}, \bibinfo {author}
  {\bibfnamefont {J.}~\bibnamefont {Mart\'in-Albo}}, \ and\ \bibinfo {author}
  {\bibfnamefont {J.}~\bibnamefont {Burguet-Castell}},\ }\href {\doibase
  10.1103/PhysRevD.86.093015} {\bibfield  {journal} {\bibinfo  {journal} {Phys.
  Rev. D}\ }\textbf {\bibinfo {volume} {86}},\ \bibinfo {pages} {093015}
  (\bibinfo {year} {2012})},\ \Eprint {http://arxiv.org/abs/1208.2735}
  {arXiv:1208.2735 [hep-ex]} \BibitemShut {NoStop}%
\bibitem [{\citenamefont {Ambrosio}\ \emph {et~al.}(2001)\citenamefont
  {Ambrosio} \emph {et~al.}}]{Ambrosio:2001ej}%
  \BibitemOpen
  \bibfield  {author} {\bibinfo {author} {\bibfnamefont {G.}~\bibnamefont
  {Ambrosio}} \emph {et~al.} (\bibinfo {collaboration} {VLHC Design Study
  Group}),\ }\href@noop {} {\emph {\bibinfo {title} {Design study for a staged
  very large hadron collider}}},\ \bibinfo {type} {Tech. Rep.}\ \bibinfo
  {number} {SLAC-R-591; FERMILAB-TM-2149}\ (\bibinfo {year} {2001})\BibitemShut
  {NoStop}%
\bibitem [{\citenamefont {{Foster, G.W. and Kashikhin, V.S. and Malamud, E. and
  Mazur, P. and Oleck, A. and Piekarz, H. and Fuerst, J. and Rabehl, R. and
  Schlabach, P. and Volk, J.}}(2000)}]{828238}%
  \BibitemOpen
  \bibfield  {author} {\bibinfo {author} {\bibnamefont {{Foster, G.W. and
  Kashikhin, V.S. and Malamud, E. and Mazur, P. and Oleck, A. and Piekarz, H.
  and Fuerst, J. and Rabehl, R. and Schlabach, P. and Volk, J.}}},\ }\href
  {\doibase 10.1109/77.828238} {\bibfield  {journal} {\bibinfo  {journal} {IEEE
  Transactions on Applied Superconductivity}\ }\textbf {\bibinfo {volume}
  {10}},\ \bibinfo {pages} {318 } (\bibinfo {year} {2000})}\BibitemShut
  {NoStop}%
\bibitem [{\citenamefont {Andreopoulos}\ \emph {et~al.}(2010)\citenamefont
  {Andreopoulos} \emph {et~al.}}]{Andreopoulos:2009rq}%
  \BibitemOpen
  \bibfield  {author} {\bibinfo {author} {\bibfnamefont {C.}~\bibnamefont
  {Andreopoulos}} \emph {et~al.},\ }\href {\doibase 10.1016/j.nima.2009.12.009}
  {\bibfield  {journal} {\bibinfo  {journal} {Nucl.Instrum.Meth.}\ }\textbf
  {\bibinfo {volume} {A614}},\ \bibinfo {pages} {87} (\bibinfo {year}
  {2010})},\ \Eprint {http://arxiv.org/abs/0905.2517} {arXiv:0905.2517
  [hep-ph]} \BibitemShut {NoStop}%
\bibitem [{\citenamefont {Cervera}\ \emph {et~al.}(2010)\citenamefont {Cervera}
  \emph {et~al.}}]{Cervera:2010rz}%
  \BibitemOpen
  \bibfield  {author} {\bibinfo {author} {\bibfnamefont {A.}~\bibnamefont
  {Cervera}} \emph {et~al.},\ }\href {\doibase 10.1016/j.nima.2010.09.049}
  {\bibfield  {journal} {\bibinfo  {journal} {Nucl. Instrum. Meth.}\ }\textbf
  {\bibinfo {volume} {A624}},\ \bibinfo {pages} {601} (\bibinfo {year}
  {2010})},\ \Eprint {http://arxiv.org/abs/1004.0358} {arXiv:1004.0358
  [hep-ex]} \BibitemShut {NoStop}%
\bibitem [{\citenamefont {Ingelman}\ \emph {et~al.}(1997)\citenamefont
  {Ingelman} \emph {et~al.}}]{Ingelman:1996mq}%
  \BibitemOpen
  \bibfield  {author} {\bibinfo {author} {\bibfnamefont {G.}~\bibnamefont
  {Ingelman}} \emph {et~al.},\ }\href {\doibase 10.1016/S0010-4655(96)00157-9}
  {\bibfield  {journal} {\bibinfo  {journal} {Comput.Phys.Commun.}\ }\textbf
  {\bibinfo {volume} {101}},\ \bibinfo {pages} {108} (\bibinfo {year}
  {1997})},\ \Eprint {http://arxiv.org/abs/hep-ph/9605286}
  {arXiv:hep-ph/9605286 [hep-ph]} \BibitemShut {NoStop}%
\bibitem [{\citenamefont {Casper}(2002)}]{Casper:2002sd}%
  \BibitemOpen
  \bibfield  {author} {\bibinfo {author} {\bibfnamefont {D.}~\bibnamefont
  {Casper}},\ }\href {\doibase 10.1016/S0920-5632(02)01756-5} {\bibfield
  {journal} {\bibinfo  {journal} {Nucl.Phys.Proc.Suppl.}\ }\textbf {\bibinfo
  {volume} {112}},\ \bibinfo {pages} {161} (\bibinfo {year} {2002})},\ \Eprint
  {http://arxiv.org/abs/hep-ph/0208030} {arXiv:hep-ph/0208030 [hep-ph]}
  \BibitemShut {NoStop}%
\bibitem [{\citenamefont {Allison}\ \emph {et~al.}(2006)\citenamefont {Allison}
  \emph {et~al.}}]{1610988}%
  \BibitemOpen
  \bibfield  {author} {\bibinfo {author} {\bibfnamefont {J.}~\bibnamefont
  {Allison}} \emph {et~al.},\ }\href {\doibase 10.1109/TNS.2006.869826}
  {\bibfield  {journal} {\bibinfo  {journal} {IEEE Transactions on Nuclear
  Science}\ }\textbf {\bibinfo {volume} {53}},\ \bibinfo {pages} {270 }
  (\bibinfo {year} {2006})}\BibitemShut {NoStop}%
\bibitem [{\citenamefont {Cervera-Villanueva}\ \emph
  {et~al.}(2004)\citenamefont {Cervera-Villanueva} \emph
  {et~al.}}]{CerveraVillanueva:2004kt}%
  \BibitemOpen
  \bibfield  {author} {\bibinfo {author} {\bibfnamefont {A.}~\bibnamefont
  {Cervera-Villanueva}} \emph {et~al.},\ }\href {\doibase
  10.1016/j.nima.2004.07.074} {\bibfield  {journal} {\bibinfo  {journal}
  {Nucl.Instrum.Meth.}\ }\textbf {\bibinfo {volume} {A534}},\ \bibinfo {pages}
  {180} (\bibinfo {year} {2004})}\BibitemShut {NoStop}%
\bibitem [{\citenamefont {Abt}\ \emph {et~al.}(2002)\citenamefont {Abt},
  \citenamefont {Emelyanov}, \citenamefont {Gorbunov},\ and\ \citenamefont
  {Kisel}}]{Abt:2002ay}%
  \BibitemOpen
  \bibfield  {author} {\bibinfo {author} {\bibfnamefont {I.}~\bibnamefont
  {Abt}}, \bibinfo {author} {\bibfnamefont {D.}~\bibnamefont {Emelyanov}},
  \bibinfo {author} {\bibfnamefont {I.}~\bibnamefont {Gorbunov}}, \ and\
  \bibinfo {author} {\bibfnamefont {I.}~\bibnamefont {Kisel}},\ }\href
  {\doibase 10.1016/S0168-9002(02)01097-5} {\bibfield  {journal} {\bibinfo
  {journal} {Nucl.Instrum.Meth.}\ }\textbf {\bibinfo {volume} {A490}},\
  \bibinfo {pages} {546} (\bibinfo {year} {2002})}\BibitemShut {NoStop}%
\bibitem [{\citenamefont {Groom}\ \emph {et~al.}(2001)\citenamefont {Groom},
  \citenamefont {Mikhov},\ and\ \citenamefont {Striganov}}]{groom:2001ms}%
  \BibitemOpen
  \bibfield  {author} {\bibinfo {author} {\bibfnamefont {D.}~\bibnamefont
  {Groom}}, \bibinfo {author} {\bibfnamefont {N.}~\bibnamefont {Mikhov}}, \
  and\ \bibinfo {author} {\bibfnamefont {S.}~\bibnamefont {Striganov}},\
  }\href@noop {} {\bibfield  {journal} {\bibinfo  {journal} {Atomic Data and
  Nuclear Data Tables}\ }\textbf {\bibinfo {volume} {78}},\ \bibinfo {pages}
  {183} (\bibinfo {year} {2001})}\BibitemShut {NoStop}%
\bibitem [{\citenamefont {Bari}\ \emph {et~al.}(2003)\citenamefont {Bari},
  \citenamefont {Candela}, \citenamefont {De~Deo}, \citenamefont {D'Incecco},
  \citenamefont {Garbini} \emph {et~al.}}]{Bari:2003bt}%
  \BibitemOpen
  \bibfield  {author} {\bibinfo {author} {\bibfnamefont {G.}~\bibnamefont
  {Bari}}, \bibinfo {author} {\bibfnamefont {A.}~\bibnamefont {Candela}},
  \bibinfo {author} {\bibfnamefont {M.}~\bibnamefont {De~Deo}}, \bibinfo
  {author} {\bibfnamefont {M.}~\bibnamefont {D'Incecco}}, \bibinfo {author}
  {\bibfnamefont {M.}~\bibnamefont {Garbini}},  \emph {et~al.},\ }\href
  {\doibase 10.1016/S0168-9002(03)01345-7} {\bibfield  {journal} {\bibinfo
  {journal} {Nucl.Instrum.Meth.}\ }\textbf {\bibinfo {volume} {A508}},\
  \bibinfo {pages} {170} (\bibinfo {year} {2003})}\BibitemShut {NoStop}%
\bibitem [{\citenamefont {Adamson}\ \emph
  {et~al.}(2011{\natexlab{b}})\citenamefont {Adamson} \emph
  {et~al.}}]{Adamson:2011fa}%
  \BibitemOpen
  \bibfield  {author} {\bibinfo {author} {\bibfnamefont {P.}~\bibnamefont
  {Adamson}} \emph {et~al.} (\bibinfo {collaboration} {MINOS collaboration}),\
  }\href {\doibase 10.1103/PhysRevLett.107.021801} {\bibfield  {journal}
  {\bibinfo  {journal} {Phys.Rev.Lett.}\ }\textbf {\bibinfo {volume} {107}},\
  \bibinfo {pages} {021801} (\bibinfo {year} {2011}{\natexlab{b}})},\ \Eprint
  {http://arxiv.org/abs/1104.0344} {arXiv:1104.0344 [hep-ex]} \BibitemShut
  {NoStop}%
\bibitem [{\citenamefont {Burguet-Castell}\ \emph {et~al.}(2001)\citenamefont
  {Burguet-Castell}, \citenamefont {Gavela}, \citenamefont {G\'omez-Cadenas},
  \citenamefont {Hern\'andez},\ and\ \citenamefont
  {Mena}}]{BurguetCastell:2001ez}%
  \BibitemOpen
  \bibfield  {author} {\bibinfo {author} {\bibfnamefont {J.}~\bibnamefont
  {Burguet-Castell}}, \bibinfo {author} {\bibfnamefont {M.~B.}\ \bibnamefont
  {Gavela}}, \bibinfo {author} {\bibfnamefont {J.~J.}\ \bibnamefont
  {G\'omez-Cadenas}}, \bibinfo {author} {\bibfnamefont {P.}~\bibnamefont
  {Hern\'andez}}, \ and\ \bibinfo {author} {\bibfnamefont {O.}~\bibnamefont
  {Mena}},\ }\href {\doibase 10.1016/S0550-3213(01)00248-6} {\bibfield
  {journal} {\bibinfo  {journal} {Nucl. Phys.}\ }\textbf {\bibinfo {volume}
  {B608}},\ \bibinfo {pages} {301} (\bibinfo {year} {2001})},\ \Eprint
  {http://arxiv.org/abs/hep-ph/0103258} {arXiv:hep-ph/0103258} \BibitemShut
  {NoStop}%
\bibitem [{\citenamefont {Burguet-Castell}\ \emph {et~al.}(2002)\citenamefont
  {Burguet-Castell}, \citenamefont {Gavela}, \citenamefont {G\'omez-Cadenas},
  \citenamefont {Hern\'andez},\ and\ \citenamefont
  {Mena}}]{BurguetCastell:2002qx}%
  \BibitemOpen
  \bibfield  {author} {\bibinfo {author} {\bibfnamefont {J.}~\bibnamefont
  {Burguet-Castell}}, \bibinfo {author} {\bibfnamefont {M.~B.}\ \bibnamefont
  {Gavela}}, \bibinfo {author} {\bibfnamefont {J.~J.}\ \bibnamefont
  {G\'omez-Cadenas}}, \bibinfo {author} {\bibfnamefont {P.}~\bibnamefont
  {Hern\'andez}}, \ and\ \bibinfo {author} {\bibfnamefont {O.}~\bibnamefont
  {Mena}},\ }\href {\doibase 10.1016/S0550-3213(02)00872-6} {\bibfield
  {journal} {\bibinfo  {journal} {Nucl. Phys.}\ }\textbf {\bibinfo {volume}
  {B646}},\ \bibinfo {pages} {301} (\bibinfo {year} {2002})},\ \Eprint
  {http://arxiv.org/abs/hep-ph/0207080} {arXiv:hep-ph/0207080} \BibitemShut
  {NoStop}%
\bibitem [{\citenamefont {Burguet-Castell}\ \emph {et~al.}(2005)\citenamefont
  {Burguet-Castell}, \citenamefont {Casper}, \citenamefont {Couce},
  \citenamefont {G\'omez-Cadenas},\ and\ \citenamefont
  {Hern\'andez}}]{BurguetCastell:2005pa}%
  \BibitemOpen
  \bibfield  {author} {\bibinfo {author} {\bibfnamefont {J.}~\bibnamefont
  {Burguet-Castell}}, \bibinfo {author} {\bibfnamefont {D.}~\bibnamefont
  {Casper}}, \bibinfo {author} {\bibfnamefont {E.}~\bibnamefont {Couce}},
  \bibinfo {author} {\bibfnamefont {J.~J.}\ \bibnamefont {G\'omez-Cadenas}}, \
  and\ \bibinfo {author} {\bibfnamefont {P.}~\bibnamefont {Hern\'andez}},\
  }\href {\doibase 10.1016/j.nuclphysb.2005.06.037} {\bibfield  {journal}
  {\bibinfo  {journal} {Nucl. Phys.}\ }\textbf {\bibinfo {volume} {B725}},\
  \bibinfo {pages} {306} (\bibinfo {year} {2005})},\ \Eprint
  {http://arxiv.org/abs/hep-ph/0503021} {arXiv:hep-ph/0503021} \BibitemShut
  {NoStop}%
\bibitem [{\citenamefont {Bogomilov}\ \emph {et~al.}(2013)\citenamefont
  {Bogomilov}, \citenamefont {Karadzhov}, \citenamefont {Matev}, \citenamefont
  {Tsenov}, \citenamefont {Laing},\ and\ \citenamefont
  {Soler}}]{NearDetector:2013}%
  \BibitemOpen
  \bibfield  {author} {\bibinfo {author} {\bibfnamefont {M.}~\bibnamefont
  {Bogomilov}}, \bibinfo {author} {\bibfnamefont {Y.}~\bibnamefont
  {Karadzhov}}, \bibinfo {author} {\bibfnamefont {R.}~\bibnamefont {Matev}},
  \bibinfo {author} {\bibfnamefont {R.}~\bibnamefont {Tsenov}}, \bibinfo
  {author} {\bibfnamefont {A.}~\bibnamefont {Laing}}, \ and\ \bibinfo {author}
  {\bibfnamefont {F.~J.~P.}\ \bibnamefont {Soler}},\ }\href@noop {} {\bibfield
  {journal} {\bibinfo  {journal} {Phys. Rev. ST Accel. Beams}\ } (\bibinfo
  {year} {2013})},\ \bibinfo {note} {accepted for publication}\BibitemShut
  {NoStop}%
\end{thebibliography}%

\section{Appendix} 
This appendix summarizes the response matrices of the wrong sign muon signal from $\nu_{\mu}$ and $\bar{\nu}_{\mu}$ appearance and the associated backgrounds in bins of true and reconstructed neutrino energy relevant to an oscillation analysis. Each entry in the table is the survival probability for each species. In all tables, columns represent the true neutrino energy in GeV and rows the reconstructed energy, also in GeV. The overflow bin in reconstructed energy represents all events with a reconstructed energy greater than the known maximum. Migration matrices assuming a negative charge focussing magnetic field and a positive charge focussing magnetic field are shown. The backgrounds generated by $\nu_{\mu}(\bar{\nu}_{\mu})$ NC interactions are consistent with zero at all energies for the 3$\times10^{6}$ events simulated. Therefore these matrices are not shown.


\begin{table}[h!] 
\subsection{$\nu_{\mu}$ Appearance Matrices, Positive Focussing Detector Field}
\begin{tabular}{|c||c|c|c|c|c|c|c|c|c|c|c|c|c|}
\hline
 & 0.0-2.0 & 2.0-2.5 & 2.5-3.0 & 3.0-3.5 & 3.5-4.0 & 4.0-4.5 & 4.5-5.0 & 5.0-5.5 & 5.5-6.0 & 6.0-7.0 & 7.0-8.0 & 8.0-9.0 & 9.0-10.0 \\
 \hline 
0.0-2.0 & 1020 & 906.9 & 458.4 & 213.8 & 120.2 & 80.49 & 52.00 & 31.51 & 23.13 & 12.37 & 10.26 & 9.963 & 6.317 \\
2.0-2.5 & 421.2 & 910.5 & 605.9 & 318.6 & 147.8 & 85.26 & 56.23 & 35.77 & 21.13 & 10.58 & 6.357 & 3.919 & 3.526 \\
2.5-3.0 & 202.5 & 863.7 & 1010 & 638.6 & 335.0 & 178.2 & 97.94 & 61.32 & 38.83 & 19.85 & 10.85 & 6.708 & 5.876 \\
3.0-3.5 & 50.25 & 482.9 & 1007 & 948.1 & 622.2 & 345.5 & 190.9 & 110.0 & 70.52 & 36.37 & 15.75 & 9.830 & 8.080 \\
3.5-4.0 & 41.38 & 229.4 & 687.9 & 1036 & 885.6 & 582.0 & 342.4 & 197.4 & 116.8 & 58.51 & 26.94 & 15.61 & 12.49 \\
4.0-4.5 & 11.82 & 115.3 & 367.8 & 792.9 & 1058 & 881.1 & 601.2 & 366.6 & 209.7 & 106.3 & 40.07 & 22.52 & 14.69 \\
4.5-5.0 & 11.82 & 30.03 & 141.9 & 474.8 & 859.6 & 1032 & 857.5 & 585.8 & 358.2 & 179.7 & 72.78 & 30.35 & 16.45 \\
5.0-5.5 & 1.478 & 14.41 & 50.61 & 198.1 & 548.2 & 870.8 & 956.9 & 805.1 & 557.6 & 289.8 & 116.0 & 52.74 & 27.03 \\
5.5-6.0 & 4.434 & 6.006 & 21.39 & 88.10 & 276.1 & 598.6 & 876.9 & 932.8 & 789.9 & 452.0 & 199.6 & 82.69 & 38.34 \\
6.0-7.0 & 2.956 & 12.01 & 22.10 & 50.95 & 164.7 & 459.1 & 999.4 & 1505 & 1735 & 1409 & 756.2 & 346.0 & 168.1 \\
7.0-8.0 & 2.956 & 7.207 & 14.97 & 14.29 & 31.65 & 87.64 & 250.5 & 602.2 & 1103 & 1566 & 1321 & 759.0 & 403.8 \\
8.0-9.0 & 0 & 3.604 & 10.69 & 10.48 & 16.50 & 25.15 & 56.43 & 148.4 & 354.9 & 936.4 & 1448 & 1237 & 815.0 \\
9.0-10.0 & 1.478 & 2.402 & 4.990 & 6.667 & 9.091 & 12.18 & 16.32 & 35.94 & 81.23 & 347.7 & 986.0 & 1355 & 1208 \\
10.0-11.0 & 4.434 & 3.604 & 9.267 & 14.29 & 15.49 & 19.59 & 29.42 & 41.39 & 57.53 & 154.4 & 625.8 & 1728 & 2973 \\
\hline
\end{tabular}
\caption{Golden channel $\nu_{\mu}$ appearance signal efficiency. All values $\times 10^{-4}$.}
\end{table}

\begin{table}[h!] 
\begin{tabular}{|c||c|c|c|c|c|c|c|c|c|c|c|c|c|}
\hline
 & 0.0-2.0 & 2.0-2.5 & 2.5-3.0 & 3.0-3.5 & 3.5-4.0 & 4.0-4.5 & 4.5-5.0 & 5.0-5.5 & 5.5-6.0 & 6.0-7.0 & 7.0-8.0 & 8.0-9.0 & 9.0-10.0 \\
 \hline 
0.0-2.0 & 4.903 & 0 & 2.659 & 1.323 & 2.980 & 1.165 & 2.110 & 1.480 & 0.8671 & 1.432 & 0.8879 & 0.5123 & 0.9993 \\
2.0-2.5 & 0 & 1.148 & 1.995 & 0 & 0.5960 & 0 & 1.407 & 0.7400 & 0.2477 & 0.5834 & 0.3946 & 0.1708 & 0.6246 \\
2.5-3.0 & 0 & 0 & 0.6648 & 0 & 0.5960 & 0.2331 & 0.1758 & 0.5920 & 0.7432 & 0.5834 & 0.5920 & 0.3985 & 0.7495 \\
3.0-3.5 & 3.269 & 0 & 0 & 1.765 & 1.192 & 1.398 & 1.231 & 1.628 & 0.6194 & 0.3182 & 0.4933 & 0.5123 & 0.6246 \\
3.5-4.0 & 0 & 1.148 & 0.6648 & 2.206 & 0.5960 & 1.398 & 0.3517 & 0.7400 & 0.8671 & 0.5834 & 0.3946 & 0.3985 & 1.124 \\
4.0-4.5 & 0 & 1.148 & 1.330 & 0.8823 & 0.5960 & 1.398 & 1.758 & 1.184 & 0.7432 & 0.4774 & 0.4933 & 0.5123 & 0.6246 \\
4.5-5.0 & 1.634 & 1.148 & 1.330 & 0.4412 & 0.8939 & 1.632 & 0.3517 & 1.036 & 0.3716 & 1.114 & 0.3453 & 0.6262 & 1.249 \\
5.0-5.5 & 0 & 0 & 1.330 & 0.4412 & 1.490 & 0.9323 & 0.5275 & 0.8879 & 1.239 & 0.5834 & 0.2960 & 1.025 & 0.2498 \\
5.5-6.0 & 0 & 0 & 1.995 & 0.8823 & 0.2980 & 0.9323 & 0 & 1.332 & 1.486 & 0.6895 & 0.8386 & 0.6831 & 0.3747 \\
6.0-7.0 & 0 & 1.148 & 0 & 2.647 & 0.8939 & 3.030 & 2.286 & 1.628 & 2.477 & 1.963 & 1.381 & 1.366 & 1.749 \\
7.0-8.0 & 0 & 1.148 & 1.330 & 1.323 & 1.192 & 1.398 & 1.758 & 2.664 & 1.734 & 1.856 & 1.677 & 1.082 & 1.624 \\
8.0-9.0 & 0 & 0 & 1.330 & 0 & 1.192 & 0.9323 & 0.8792 & 1.776 & 1.486 & 1.750 & 1.529 & 1.423 & 1.499 \\
9.0-10.0 & 1.634 & 1.148 & 1.330 & 0 & 1.490 & 1.632 & 0.5275 & 1.628 & 1.363 & 1.432 & 1.135 & 1.537 & 1.499 \\
10.0-11.0 & 0 & 3.445 & 0 & 0.8823 & 1.490 & 2.098 & 1.758 & 2.516 & 2.849 & 3.501 & 4.094 & 4.554 & 4.997 \\
\hline
\end{tabular}
\caption{$\mu^{-}$ background from charge mis-identified $\bar{\nu}_{\mu}$ CC events All values $\times 10^{-4}$.}
\end{table}

\begin{table}[h!] 
\begin{tabular}{|c||c|c|c|c|c|c|c|c|c|c|c|c|c|}
\hline
 & 0.0-2.0 & 2.0-2.5 & 2.5-3.0 & 3.0-3.5 & 3.5-4.0 & 4.0-4.5 & 4.5-5.0 & 5.0-5.5 & 5.5-6.0 & 6.0-7.0 & 7.0-8.0 & 8.0-9.0 & 9.0-10.0 \\
 \hline 
0.0-2.0 & 0 & 0 & 0.3245 & 0.2196 & 0 & 0.1185 & 0.0919 & 0.0775 & 0.2626 & 0.1125 & 0.0528 & 0.1824 & 0.2014 \\
2.0-2.5 & 0 & 0 & 0.3245 & 0.2196 & 0.3047 & 0.1185 & 0.0919 & 0.0775 & 0 & 0.0562 & 0.0264 & 0.0608 & 0 \\
2.5-3.0 & 0 & 0 & 0 & 0.2196 & 0 & 0 & 0 & 0.1550 & 0.0657 & 0.0562 & 0.0264 & 0 & 0.0671 \\
3.0-3.5 & 0 & 0 & 0.3245 & 0.2196 & 0.1524 & 0 & 0 & 0.1550 & 0.2626 & 0.1125 & 0.0792 & 0.1216 & 0.0671 \\
3.5-4.0 & 0 & 0 & 0 & 0 & 0 & 0.2370 & 0.0919 & 0.1550 & 0.1970 & 0.0562 & 0.2642 & 0.1216 & 0.2014 \\
4.0-4.5 & 0 & 0 & 0 & 0.2196 & 0.1524 & 0.1185 & 0.3675 & 0.3876 & 0.5909 & 0.3375 & 0.2377 & 0.0912 & 0.2686 \\
4.5-5.0 & 0 & 0 & 0 & 0 & 0.3047 & 0.1185 & 0.1838 & 0.3101 & 0.3283 & 0.4218 & 0.4226 & 0.3648 & 0.3357 \\
5.0-5.5 & 0 & 0 & 0 & 0.4393 & 0.1524 & 0.5924 & 0.5513 & 0.6201 & 0.1313 & 0.4218 & 0.3962 & 0.4864 & 0.1343 \\
5.5-6.0 & 0 & 0 & 0 & 0.2196 & 0.1524 & 0.5924 & 0.7351 & 0.7752 & 0.6565 & 0.7593 & 0.4491 & 0.3952 & 0.7386 \\
6.0-7.0 & 0 & 0 & 0 & 0.6589 & 0 & 0.8294 & 1.195 & 1.395 & 0.9191 & 1.434 & 1.321 & 1.337 & 1.276 \\
7.0-8.0 & 0 & 0 & 0 & 0.2196 & 0.4571 & 0.3555 & 0.2757 & 1.008 & 1.247 & 1.294 & 1.559 & 1.733 & 1.343 \\
8.0-9.0 & 0 & 0 & 0 & 0.2196 & 0.4571 & 0.4740 & 0.1838 & 0.6976 & 1.247 & 1.350 & 1.321 & 1.520 & 1.343 \\
9.0-10.0 & 0 & 0 & 0.3245 & 0 & 0 & 0.2370 & 0.0919 & 0.4651 & 0.4596 & 0.9562 & 1.506 & 1.216 & 1.343 \\
10.0-11.0 & 0 & 0 & 0 & 0 & 0 & 0.2370 & 0.3675 & 1.008 & 0.8535 & 1.828 & 2.932 & 4.104 & 5.909 \\
\hline
\end{tabular}
\caption{$\mu^{-}$ background from $\nu_{e}$ CC events. All values $\times 10^{-4}$.}
\end{table}

\begin{table}[h!] 
\begin{tabular}{|c||c|c|c|c|c|c|c|c|c|c|c|c|c|}
\hline
 & 0.0-2.0 & 2.0-2.5 & 2.5-3.0 & 3.0-3.5 & 3.5-4.0 & 4.0-4.5 & 4.5-5.0 & 5.0-5.5 & 5.5-6.0 & 6.0-7.0 & 7.0-8.0 & 8.0-9.0 & 9.0-10.0 \\
 \hline 
0.0-2.0 & 0 & 0 & 0 & 128.2 & 100.00 & 95.54 & 72.60 & 78.94 & 76.18 & 70.46 & 61.40 & 49.75 & 41.53 \\
2.0-2.5 & 0 & 0 & 0 & 0 & 88.68 & 75.13 & 65.80 & 65.66 & 62.45 & 59.56 & 49.96 & 41.02 & 32.78 \\
2.5-3.0 & 0 & 0 & 0 & 128.2 & 81.13 & 87.38 & 82.06 & 85.33 & 81.54 & 77.24 & 64.96 & 57.31 & 50.62 \\
3.0-3.5 & 0 & 0 & 0 & 0 & 60.38 & 93.91 & 104.4 & 98.61 & 95.94 & 90.35 & 77.32 & 66.57 & 58.34 \\
3.5-4.0 & 0 & 0 & 0 & 0 & 67.92 & 69.41 & 85.46 & 92.46 & 94.10 & 93.35 & 86.64 & 78.91 & 64.03 \\
4.0-4.5 & 0 & 0 & 0 & 0 & 30.19 & 52.26 & 71.09 & 84.35 & 85.89 & 90.75 & 86.34 & 78.00 & 68.37 \\
4.5-5.0 & 0 & 0 & 0 & 0 & 18.87 & 39.20 & 44.62 & 65.17 & 72.50 & 81.37 & 85.21 & 78.30 & 76.60 \\
5.0-5.5 & 0 & 0 & 0 & 0 & 5.660 & 12.25 & 23.07 & 42.79 & 55.92 & 72.21 & 76.45 & 74.74 & 70.23 \\
5.5-6.0 & 0 & 0 & 0 & 0 & 1.887 & 9.799 & 16.64 & 26.31 & 39.68 & 54.64 & 68.86 & 71.61 & 72.95 \\
6.0-7.0 & 0 & 0 & 0 & 0 & 1.887 & 7.349 & 12.10 & 22.87 & 37.67 & 68.09 & 102.7 & 127.1 & 129.6 \\
7.0-8.0 & 0 & 0 & 0 & 0 & 0 & 2.450 & 3.781 & 4.918 & 11.22 & 28.03 & 58.76 & 88.69 & 99.62 \\
8.0-9.0 & 0 & 0 & 0 & 0 & 0 & 0.8166 & 0.7563 & 2.459 & 3.851 & 9.154 & 26.24 & 51.84 & 74.56 \\
9.0-10.0 & 0 & 0 & 0 & 0 & 0 & 0 & 0.3781 & 0.7377 & 1.172 & 2.712 & 9.844 & 22.99 & 44.42 \\
10.0-11.0 & 0 & 0 & 0 & 0 & 1.887 & 1.633 & 3.025 & 2.951 & 2.512 & 3.164 & 6.028 & 15.56 & 35.84 \\
\hline
\end{tabular}
\caption{$\mu^{-}$ reconstructed from $\nu_{\tau}$ CC events. All values $\times 10^{-4}$.}
\end{table}

\begin{table}[h!] 
\begin{tabular}{|c||c|c|c|c|c|c|c|c|c|c|c|c|c|}
\hline
 & 0.0-2.0 & 2.0-2.5 & 2.5-3.0 & 3.0-3.5 & 3.5-4.0 & 4.0-4.5 & 4.5-5.0 & 5.0-5.5 & 5.5-6.0 & 6.0-7.0 & 7.0-8.0 & 8.0-9.0 & 9.0-10.0 \\
 \hline 
0.0-2.0 & 0 & 0 & 0 & 0 & 0 & 0 & 0.8818 & 0.4853 & 1.102 & 1.350 & 1.851 & 1.887 & 1.658 \\
2.0-2.5 & 0 & 0 & 0 & 0 & 0 & 0 & 0 & 0.7279 & 0.6295 & 0.6501 & 0.9820 & 1.132 & 1.357 \\
2.5-3.0 & 0 & 0 & 0 & 0 & 0 & 0 & 0.4409 & 0.7279 & 0.9443 & 0.8001 & 1.020 & 1.283 & 1.131 \\
3.0-3.5 & 0 & 0 & 0 & 0 & 0 & 0 & 0 & 0.4853 & 0.9443 & 1.000 & 0.9820 & 0.9813 & 1.507 \\
3.5-4.0 & 0 & 0 & 0 & 0 & 0 & 1.213 & 0.4409 & 0.4853 & 0.7869 & 0.8501 & 0.9065 & 1.547 & 1.357 \\
4.0-4.5 & 0 & 0 & 0 & 0 & 0 & 0 & 0 & 0.7279 & 0.4721 & 0.8001 & 1.095 & 1.094 & 1.357 \\
4.5-5.0 & 0 & 0 & 0 & 0 & 0 & 1.213 & 0 & 0 & 0.3148 & 0.4001 & 0.5666 & 0.7926 & 1.281 \\
5.0-5.5 & 0 & 0 & 0 & 0 & 0 & 0 & 0 & 0.7279 & 0.1574 & 0.3501 & 0.5288 & 0.4152 & 1.055 \\
5.5-6.0 & 0 & 0 & 0 & 0 & 0 & 0 & 0.8818 & 0.2426 & 0.3148 & 0.1500 & 0.5288 & 0.5661 & 0.7537 \\
6.0-7.0 & 0 & 0 & 0 & 0 & 0 & 0 & 0 & 0 & 0.1574 & 0.2500 & 0.4533 & 0.9435 & 1.131 \\
7.0-8.0 & 0 & 0 & 0 & 0 & 0 & 0 & 0 & 0 & 0 & 0.3501 & 0.2644 & 0.6793 & 1.131 \\
8.0-9.0 & 0 & 0 & 0 & 0 & 0 & 0 & 0 & 0 & 0 & 0.1500 & 0.3399 & 0.1887 & 0.3015 \\
9.0-10.0 & 0 & 0 & 0 & 0 & 0 & 0 & 0 & 0 & 0 & 0.1500 & 0.1889 & 0.1510 & 0.2261 \\
10.0-11.0 & 0 & 0 & 0 & 0 & 0 & 0 & 0 & 0 & 0.1574 & 0.3000 & 0.1511 & 0.3774 & 0.6030 \\
\hline
\end{tabular}
\caption{$\mu^{-}$ background from $\bar{\nu}_{\tau}$ CC events. All values $\times 10^{-4}$.}
\end{table}

\begin{table}[h!] 
\subsection{$\bar{\nu}_{\mu}$ Appearance Matrices, Positive Focussing Detector Field}
\begin{tabular}{|c||c|c|c|c|c|c|c|c|c|c|c|c|c|}
\hline
 & 0.0-2.0 & 2.0-2.5 & 2.5-3.0 & 3.0-3.5 & 3.5-4.0 & 4.0-4.5 & 4.5-5.0 & 5.0-5.5 & 5.5-6.0 & 6.0-7.0 & 7.0-8.0 & 8.0-9.0 & 9.0-10.0 \\
 \hline 
0.0-2.0 & 1173 & 1326 & 688.8 & 296.0 & 156.7 & 74.59 & 41.50 & 23.97 & 20.69 & 13.37 & 11.00 & 10.08 & 8.994 \\
2.0-2.5 & 590.0 & 1282 & 865.0 & 448.7 & 199.4 & 99.76 & 46.78 & 28.86 & 20.81 & 13.10 & 8.978 & 5.920 & 4.622 \\
2.5-3.0 & 223.9 & 1263 & 1373 & 896.9 & 449.7 & 198.8 & 95.31 & 54.76 & 33.32 & 19.41 & 12.28 & 10.47 & 9.368 \\
3.0-3.5 & 65.37 & 666.1 & 1443 & 1364 & 832.6 & 426.3 & 210.0 & 112.2 & 64.41 & 33.73 & 18.79 & 13.95 & 9.743 \\
3.5-4.0 & 29.42 & 259.5 & 979.3 & 1515 & 1348 & 799.7 & 412.0 & 217.8 & 119.4 & 52.25 & 27.53 & 17.99 & 17.36 \\
4.0-4.5 & 16.34 & 95.31 & 452.1 & 1147 & 1517 & 1211 & 768.1 & 431.4 & 223.5 & 92.93 & 39.96 & 23.68 & 18.24 \\
4.5-5.0 & 3.269 & 33.30 & 182.8 & 615.4 & 1246 & 1504 & 1201 & 732.7 & 404.3 & 179.0 & 65.51 & 35.01 & 22.98 \\
5.0-5.5 & 3.269 & 14.93 & 76.46 & 274.4 & 723.5 & 1270 & 1426 & 1127 & 718.2 & 319.1 & 111.0 & 49.07 & 30.73 \\
5.5-6.0 & 9.806 & 21.82 & 33.24 & 101.0 & 346.3 & 827.9 & 1291 & 1341 & 1058 & 544.7 & 192.7 & 74.91 & 43.97 \\
6.0-7.0 & 14.71 & 34.45 & 41.22 & 74.55 & 203.5 & 633.5 & 1421 & 2208 & 2487 & 1907 & 842.8 & 325.9 & 146.0 \\
7.0-8.0 & 8.171 & 21.82 & 32.58 & 38.82 & 59.89 & 127.5 & 343.4 & 876.4 & 1628 & 2233 & 1728 & 849.6 & 383.9 \\
8.0-9.0 & 9.806 & 19.52 & 21.28 & 22.94 & 36.65 & 50.81 & 76.32 & 207.0 & 522.4 & 1395 & 2078 & 1596 & 893.9 \\
9.0-10.0 & 3.269 & 8.039 & 23.27 & 15.00 & 27.12 & 28.20 & 38.86 & 61.71 & 127.6 & 512.3 & 1444 & 1940 & 1579 \\
10.0-11.0 & 4.903 & 20.67 & 19.28 & 29.56 & 31.88 & 40.79 & 53.81 & 63.34 & 94.27 & 231.3 & 957.7 & 2579 & 4343 \\
\hline
\end{tabular}
\caption{Golden channel $\bar{\nu}_{\mu}$ appearance signal efficiency. All values $\times 10^{-4}$.}
\end{table}

\begin{table}[h!] 
\begin{tabular}{|c||c|c|c|c|c|c|c|c|c|c|c|c|c|}
\hline
 & 0.0-2.0 & 2.0-2.5 & 2.5-3.0 & 3.0-3.5 & 3.5-4.0 & 4.0-4.5 & 4.5-5.0 & 5.0-5.5 & 5.5-6.0 & 6.0-7.0 & 7.0-8.0 & 8.0-9.0 & 9.0-10.0 \\
 \hline 
0.0-2.0 & 1.478 & 1.201 & 2.139 & 0.9524 & 3.704 & 1.589 & 1.008 & 1.533 & 1.570 & 2.598 & 1.691 & 1.395 & 0.8814 \\
2.0-2.5 & 1.478 & 3.604 & 0 & 0.9524 & 1.347 & 0.2648 & 0.6046 & 0.8517 & 0.5710 & 0.3093 & 0.6998 & 0.6642 & 0.8814 \\
2.5-3.0 & 0 & 6.006 & 0.7129 & 0.9524 & 1.347 & 0.7943 & 0.8061 & 1.022 & 0.7138 & 0.6804 & 0.6998 & 0.8634 & 0.8814 \\
3.0-3.5 & 0 & 0 & 0 & 0 & 1.010 & 2.118 & 0.4031 & 0.5110 & 0.8566 & 0.7422 & 0.5832 & 0.6642 & 0.7345 \\
3.5-4.0 & 0 & 0 & 0 & 0.4762 & 2.020 & 0.7943 & 1.814 & 1.533 & 1.428 & 1.113 & 0.6998 & 0.7306 & 0 \\
4.0-4.5 & 0 & 0 & 1.426 & 1.429 & 0.6734 & 1.324 & 1.209 & 0.6814 & 1.285 & 0.9278 & 0.7581 & 1.063 & 0.5876 \\
4.5-5.0 & 0 & 0 & 0.7129 & 1.429 & 0.6734 & 1.853 & 0.8061 & 1.192 & 1.856 & 1.175 & 1.108 & 0.6642 & 0.4407 \\
5.0-5.5 & 0 & 2.402 & 0 & 0.4762 & 2.020 & 1.324 & 1.008 & 1.022 & 0.7138 & 1.051 & 0.9914 & 1.063 & 0.1469 \\
5.5-6.0 & 0 & 2.402 & 0.7129 & 1.429 & 0.6734 & 2.383 & 1.612 & 0.8517 & 1.428 & 1.175 & 1.050 & 1.328 & 1.028 \\
6.0-7.0 & 0 & 0 & 0 & 0.4762 & 4.041 & 3.707 & 3.225 & 2.896 & 3.569 & 3.464 & 2.566 & 2.524 & 2.350 \\
7.0-8.0 & 0 & 0 & 0 & 0.4762 & 0.6734 & 2.912 & 2.217 & 4.088 & 4.283 & 3.711 & 2.566 & 2.856 & 1.616 \\
8.0-9.0 & 0 & 0 & 0.7129 & 1.429 & 1.347 & 2.648 & 2.217 & 2.555 & 3.997 & 2.412 & 2.799 & 3.321 & 3.673 \\
9.0-10.0 & 0 & 1.201 & 0 & 0 & 2.020 & 0.7943 & 1.008 & 1.192 & 1.142 & 2.845 & 4.024 & 3.122 & 2.644 \\
10.0-11.0 & 0 & 0 & 2.139 & 1.429 & 2.020 & 1.324 & 3.628 & 3.748 & 8.423 & 7.793 & 11.26 & 11.96 & 15.28 \\
\hline
\end{tabular}
\caption{$\mu^{+}$ background from charge mis-identified $\nu_{\mu}$ CC events All values $\times 10^{-4}$.}
\end{table}

\begin{table}[h!] 
\begin{tabular}{|c||c|c|c|c|c|c|c|c|c|c|c|c|c|}
\hline
 & 0.0-2.0 & 2.0-2.5 & 2.5-3.0 & 3.0-3.5 & 3.5-4.0 & 4.0-4.5 & 4.5-5.0 & 5.0-5.5 & 5.5-6.0 & 6.0-7.0 & 7.0-8.0 & 8.0-9.0 & 9.0-10.0 \\
 \hline 
0.0-2.0 & 0 & 0.4653 & 0 & 0 & 0 & 0.0928 & 0 & 0.1196 & 0.0505 & 0 & 0.0401 & 0.0230 & 0.2038 \\
2.0-2.5 & 0 & 0 & 0 & 0 & 0 & 0 & 0 & 0 & 0 & 0 & 0.0200 & 0 & 0 \\
2.5-3.0 & 0 & 0 & 0 & 0 & 0 & 0.0928 & 0 & 0.0598 & 0 & 0.0643 & 0 & 0 & 0.0509 \\
3.0-3.5 & 0 & 0 & 0 & 0.1763 & 0.1211 & 0 & 0.0719 & 0.0598 & 0.1514 & 0.0429 & 0.0401 & 0.0230 & 0.0509 \\
3.5-4.0 & 0 & 0 & 0 & 0.1763 & 0.1211 & 0 & 0.0719 & 0.1196 & 0.1514 & 0.0429 & 0.0601 & 0.0920 & 0.0509 \\
4.0-4.5 & 0 & 0 & 0 & 0 & 0.1211 & 0.1856 & 0.0719 & 0 & 0.1514 & 0.0643 & 0.0802 & 0.1150 & 0.1019 \\
4.5-5.0 & 0 & 0 & 0 & 0.1763 & 0 & 0 & 0.2157 & 0 & 0.0505 & 0.1072 & 0.0802 & 0.0460 & 0 \\
5.0-5.5 & 0 & 0 & 0 & 0 & 0 & 0 & 0.1438 & 0 & 0.2523 & 0.1287 & 0.1403 & 0.1610 & 0.2547 \\
5.5-6.0 & 0 & 0 & 0 & 0 & 0 & 0 & 0.2157 & 0.0598 & 0.1009 & 0.1501 & 0.1403 & 0.1150 & 0.1019 \\
6.0-7.0 & 0 & 0 & 0 & 0 & 0 & 0.1856 & 0.1438 & 0.0598 & 0.2523 & 0.0858 & 0.2405 & 0.2760 & 0.1019 \\
7.0-8.0 & 0 & 0 & 0 & 0 & 0 & 0 & 0.2876 & 0.0598 & 0.0505 & 0.1930 & 0.3406 & 0.2760 & 0.5094 \\
8.0-9.0 & 0 & 0 & 0 & 0 & 0 & 0.0928 & 0.1438 & 0 & 0.1009 & 0.1501 & 0.2605 & 0.3450 & 0.4076 \\
9.0-10.0 & 0 & 0 & 0 & 0 & 0 & 0 & 0 & 0 & 0.1009 & 0.0429 & 0.1803 & 0.1380 & 0.2038 \\
10.0-11.0 & 0 & 0 & 0 & 0 & 0 & 0 & 0 & 0.0598 & 0 & 0.1716 & 0.3406 & 0.3680 & 0.8151 \\
\hline
\end{tabular}
\caption{$\mu^{+}$ background from $\bar{\nu}_{e}$ CC events. All values $\times 10^{-4}$.}
\end{table}

\begin{table}[h!] 
\begin{tabular}{|c||c|c|c|c|c|c|c|c|c|c|c|c|c|}
\hline
 & 0.0-2.0 & 2.0-2.5 & 2.5-3.0 & 3.0-3.5 & 3.5-4.0 & 4.0-4.5 & 4.5-5.0 & 5.0-5.5 & 5.5-6.0 & 6.0-7.0 & 7.0-8.0 & 8.0-9.0 & 9.0-10.0 \\
 \hline 
0.0-2.0 & 0 & 0 & 0 & 0 & 72.13 & 82.46 & 63.49 & 65.76 & 64.37 & 60.06 & 59.38 & 51.44 & 47.33 \\
2.0-2.5 & 0 & 0 & 0 & 0 & 82.43 & 81.25 & 61.28 & 66.24 & 61.22 & 60.16 & 53.82 & 46.57 & 46.13 \\
2.5-3.0 & 0 & 0 & 0 & 0 & 82.43 & 99.44 & 82.01 & 89.29 & 89.55 & 77.11 & 72.22 & 67.18 & 61.50 \\
3.0-3.5 & 0 & 0 & 0 & 0 & 77.28 & 109.1 & 88.18 & 93.17 & 93.96 & 96.22 & 85.36 & 77.52 & 71.45 \\
3.5-4.0 & 0 & 0 & 0 & 0 & 30.91 & 76.40 & 90.82 & 99.24 & 100.6 & 101.6 & 94.43 & 91.41 & 79.44 \\
4.0-4.5 & 0 & 0 & 0 & 0 & 25.76 & 50.93 & 75.39 & 89.78 & 95.37 & 100.3 & 98.05 & 93.67 & 82.91 \\
4.5-5.0 & 0 & 0 & 0 & 0 & 15.46 & 38.81 & 55.55 & 59.45 & 76.80 & 89.61 & 94.62 & 90.05 & 89.46 \\
5.0-5.5 & 0 & 0 & 0 & 0 & 10.30 & 8.489 & 28.22 & 47.07 & 66.41 & 78.76 & 89.63 & 91.11 & 84.11 \\
5.5-6.0 & 0 & 0 & 0 & 0 & 0 & 9.702 & 14.99 & 26.45 & 41.55 & 61.96 & 75.16 & 78.16 & 73.94 \\
6.0-7.0 & 0 & 0 & 0 & 0 & 0 & 8.489 & 10.14 & 23.54 & 45.80 & 72.66 & 109.0 & 132.0 & 146.4 \\
7.0-8.0 & 0 & 0 & 0 & 0 & 0 & 1.213 & 3.968 & 7.037 & 10.23 & 31.96 & 62.06 & 94.62 & 110.9 \\
8.0-9.0 & 0 & 0 & 0 & 0 & 5.152 & 3.638 & 0 & 2.184 & 2.361 & 8.551 & 29.27 & 55.33 & 78.01 \\
9.0-10.0 & 0 & 0 & 0 & 0 & 0 & 0 & 1.323 & 2.426 & 0.6295 & 3.701 & 10.46 & 25.97 & 48.84 \\
10.0-11.0 & 0 & 0 & 0 & 0 & 0 & 2.425 & 1.323 & 1.698 & 1.416 & 3.050 & 7.139 & 16.91 & 38.29 \\
\hline
\end{tabular}
\caption{$\mu^{+}$ reconstructed from $\bar{\nu}_{\tau}$ CC events. All values $\times 10^{-4}$.}
\end{table}

\begin{table}[h!] 
\begin{tabular}{|c||c|c|c|c|c|c|c|c|c|c|c|c|c|}
\hline
 & 0.0-2.0 & 2.0-2.5 & 2.5-3.0 & 3.0-3.5 & 3.5-4.0 & 4.0-4.5 & 4.5-5.0 & 5.0-5.5 & 5.5-6.0 & 6.0-7.0 & 7.0-8.0 & 8.0-9.0 & 9.0-10.0 \\
 \hline 
0.0-2.0 & 0 & 0 & 0 & 0 & 1.887 & 1.633 & 1.513 & 2.459 & 1.842 & 3.108 & 3.729 & 3.216 & 3.652 \\
2.0-2.5 & 0 & 0 & 0 & 0 & 0 & 1.633 & 1.134 & 2.951 & 1.842 & 1.469 & 2.385 & 2.651 & 2.378 \\
2.5-3.0 & 0 & 0 & 0 & 0 & 0 & 1.633 & 0.7563 & 0.9836 & 1.842 & 2.204 & 2.515 & 3.129 & 3.907 \\
3.0-3.5 & 0 & 0 & 0 & 0 & 3.774 & 1.633 & 0.3781 & 1.230 & 2.512 & 2.091 & 3.035 & 3.216 & 3.397 \\
3.5-4.0 & 0 & 0 & 0 & 0 & 0 & 0 & 0.7563 & 1.230 & 1.674 & 1.921 & 2.212 & 2.998 & 3.227 \\
4.0-4.5 & 0 & 0 & 0 & 0 & 0 & 0 & 0 & 0.4918 & 1.172 & 1.695 & 1.648 & 2.346 & 2.548 \\
4.5-5.0 & 0 & 0 & 0 & 0 & 0 & 0 & 0.3781 & 0.2459 & 1.172 & 0.9041 & 1.561 & 2.086 & 2.548 \\
5.0-5.5 & 0 & 0 & 0 & 0 & 0 & 0 & 0 & 0 & 0.8372 & 0.7911 & 1.214 & 1.564 & 1.699 \\
5.5-6.0 & 0 & 0 & 0 & 0 & 0 & 0 & 0.7563 & 0 & 0 & 0.7346 & 1.084 & 1.391 & 2.038 \\
6.0-7.0 & 0 & 0 & 0 & 0 & 1.887 & 0 & 0 & 0.4918 & 1.842 & 0.9606 & 0.9540 & 1.999 & 4.416 \\
7.0-8.0 & 0 & 0 & 0 & 0 & 0 & 0 & 0.3781 & 0.2459 & 1.005 & 0.7911 & 0.9974 & 1.651 & 2.718 \\
8.0-9.0 & 0 & 0 & 0 & 0 & 0 & 0 & 0.7563 & 0 & 0.3349 & 0.2260 & 0.4770 & 1.217 & 1.359 \\
9.0-10.0 & 0 & 0 & 0 & 0 & 0 & 0 & 0 & 0.4918 & 0.1674 & 0.0565 & 0.5637 & 0.6953 & 0.6794 \\
10.0-11.0 & 0 & 0 & 0 & 0 & 0 & 0 & 0 & 0 & 0.5023 & 0.2260 & 0.5204 & 1.173 & 2.463 \\
\hline
\end{tabular}
\caption{$\mu^{+}$ background from $\nu_{\tau}$ CC events. All values $\times 10^{-4}$.}
\end{table}


\begin{table}[h!] 
\subsection{$\nu_{\mu}$ Appearance Matrices, Negative Focussing Detector Field}
\begin{tabular}{|c||c|c|c|c|c|c|c|c|c|c|c|c|c|}
\hline
 & 0.0-2.0 & 2.0-2.5 & 2.5-3.0 & 3.0-3.5 & 3.5-4.0 & 4.0-4.5 & 4.5-5.0 & 5.0-5.5 & 5.5-6.0 & 6.0-7.0 & 7.0-8.0 & 8.0-9.0 & 9.0-10.0 \\
 \hline 
0.0-2.0 & 699.7 & 895.8 & 484.8 & 234.4 & 131.6 & 82.09 & 56.38 & 38.55 & 24.34 & 18.30 & 16.46 & 12.79 & 15.34 \\
2.0-2.5 & 293.4 & 931.0 & 609.9 & 327.4 & 173.7 & 88.46 & 57.12 & 39.49 & 26.31 & 15.23 & 10.48 & 8.203 & 6.214 \\
2.5-3.0 & 164.8 & 887.7 & 985.1 & 632.1 & 355.1 & 185.6 & 105.8 & 69.74 & 44.73 & 23.87 & 15.61 & 10.84 & 8.858 \\
3.0-3.5 & 45.82 & 577.9 & 1056 & 977.6 & 648.4 & 363.5 & 202.5 & 123.8 & 73.67 & 38.08 & 18.87 & 13.71 & 10.31 \\
3.5-4.0 & 32.83 & 264.6 & 769.5 & 1133 & 1016 & 648.2 & 367.0 & 220.5 & 130.5 & 62.57 & 30.09 & 17.94 & 11.63 \\
4.0-4.5 & 13.00 & 108.4 & 396.6 & 908.6 & 1101 & 939.1 & 648.2 & 388.8 & 233.8 & 110.4 & 49.33 & 26.38 & 16.13 \\
4.5-5.0 & 8.207 & 37.93 & 167.4 & 517.1 & 919.0 & 1114 & 919.4 & 642.5 & 402.6 & 193.6 & 77.18 & 38.50 & 19.96 \\
5.0-5.5 & 5.472 & 25.28 & 86.95 & 233.6 & 584.2 & 966.0 & 1074 & 890.6 & 612.8 & 322.0 & 129.6 & 57.48 & 30.54 \\
5.5-6.0 & 1.368 & 12.64 & 44.07 & 97.56 & 297.7 & 665.2 & 1003 & 1058 & 855.1 & 511.5 & 215.9 & 92.99 & 43.76 \\
6.0-7.0 & 7.523 & 18.96 & 26.20 & 75.23 & 198.2 & 549.6 & 1156 & 1719 & 1964 & 1582 & 834.7 & 377.6 & 176.2 \\
7.0-8.0 & 1.368 & 15.35 & 19.06 & 26.04 & 51.23 & 119.1 & 320.9 & 721.5 & 1291 & 1792 & 1493 & 852.5 & 419.5 \\
8.0-9.0 & 1.368 & 7.224 & 7.742 & 16.12 & 21.49 & 39.39 & 76.70 & 192.9 & 466.8 & 1123 & 1708 & 1420 & 882.0 \\
9.0-10.0 & 3.420 & 7.224 & 6.551 & 13.64 & 17.67 & 18.40 & 33.14 & 48.58 & 114.1 & 436.6 & 1178 & 1593 & 1401 \\
10.0-11.0 & 6.156 & 18.06 & 13.10 & 19.43 & 20.61 & 31.85 & 39.54 & 60.96 & 83.80 & 214.0 & 814.0 & 2165 & 3717 \\
\hline
\end{tabular}
\caption{Golden channel $\nu_{\mu}$ appearance signal efficiency. All values $\times 10^{-4}$.}
\end{table}

\begin{table}[h!] 
\begin{tabular}{|c||c|c|c|c|c|c|c|c|c|c|c|c|c|}
\hline
 & 0.0-2.0 & 2.0-2.5 & 2.5-3.0 & 3.0-3.5 & 3.5-4.0 & 4.0-4.5 & 4.5-5.0 & 5.0-5.5 & 5.5-6.0 & 6.0-7.0 & 7.0-8.0 & 8.0-9.0 & 9.0-10.0 \\
 \hline 
0.0-2.0 & 0 & 3.677 & 0.7089 & 1.381 & 2.192 & 2.168 & 1.666 & 1.241 & 1.972 & 1.063 & 1.143 & 1.430 & 0.7814 \\
2.0-2.5 & 1.699 & 1.226 & 2.127 & 0.4604 & 1.253 & 0 & 0.3702 & 1.086 & 0.1315 & 0.1679 & 0.4676 & 0.6554 & 0.7814 \\
2.5-3.0 & 0 & 2.451 & 1.418 & 1.381 & 0.3132 & 1.686 & 1.296 & 0.1552 & 0.6573 & 0.3357 & 0.6235 & 0.6554 & 0.5209 \\
3.0-3.5 & 1.699 & 0 & 1.418 & 1.841 & 1.253 & 0.9637 & 0.3702 & 0.3103 & 0.5258 & 0.6155 & 0.4676 & 0.5363 & 0.6511 \\
3.5-4.0 & 0 & 1.226 & 1.418 & 2.762 & 0.9395 & 0.7228 & 1.481 & 0.6206 & 0.3944 & 0.4476 & 0.7274 & 0.3575 & 0.7814 \\
4.0-4.5 & 0 & 0 & 0 & 1.841 & 2.192 & 1.205 & 1.666 & 0.7758 & 0.9202 & 0.9512 & 0.6235 & 0.6554 & 0.9116 \\
4.5-5.0 & 1.699 & 1.226 & 0.7089 & 1.841 & 1.879 & 2.409 & 1.851 & 1.241 & 0.9202 & 0.7834 & 1.091 & 0.7746 & 0.9116 \\
5.0-5.5 & 0 & 1.226 & 2.127 & 0.9207 & 1.566 & 1.686 & 2.036 & 1.552 & 1.577 & 0.3917 & 0.8833 & 1.073 & 0.6511 \\
5.5-6.0 & 1.699 & 0 & 0.7089 & 2.302 & 2.192 & 0.9637 & 1.666 & 0.9310 & 1.840 & 1.343 & 0.9872 & 0.5363 & 0.6511 \\
6.0-7.0 & 0 & 2.451 & 1.418 & 0.9207 & 4.071 & 2.891 & 2.962 & 3.258 & 3.023 & 2.350 & 2.182 & 1.966 & 2.214 \\
7.0-8.0 & 0 & 1.226 & 2.127 & 0.9207 & 1.253 & 1.446 & 2.036 & 3.569 & 3.944 & 2.350 & 2.494 & 2.026 & 2.735 \\
8.0-9.0 & 0 & 0 & 1.418 & 0.9207 & 0.9395 & 2.168 & 2.036 & 1.552 & 3.418 & 2.014 & 2.546 & 1.728 & 1.693 \\
9.0-10.0 & 0 & 1.226 & 1.418 & 1.381 & 0.3132 & 1.205 & 0.9255 & 1.862 & 1.972 & 2.966 & 2.286 & 2.264 & 2.214 \\
10.0-11.0 & 0 & 0 & 0.7089 & 1.841 & 1.566 & 2.891 & 3.517 & 3.724 & 3.549 & 5.483 & 9.457 & 10.55 & 11.33 \\
\hline
\end{tabular}
\caption{$\mu^{-}$ background from charge mis-identified $\bar{\nu}_{\mu}$ CC events All values $\times 10^{-4}$.}
\end{table}

\begin{table}[h!] 
\begin{tabular}{|c||c|c|c|c|c|c|c|c|c|c|c|c|c|}
\hline
 & 0.0-2.0 & 2.0-2.5 & 2.5-3.0 & 3.0-3.5 & 3.5-4.0 & 4.0-4.5 & 4.5-5.0 & 5.0-5.5 & 5.5-6.0 & 6.0-7.0 & 7.0-8.0 & 8.0-9.0 & 9.0-10.0 \\
 \hline 
0.0-2.0 & 0 & 0 & 0 & 52.91 & 113.9 & 105.9 & 84.95 & 87.31 & 83.50 & 74.72 & 64.56 & 56.11 & 44.50 \\
2.0-2.5 & 0 & 0 & 0 & 105.8 & 70.35 & 79.62 & 68.23 & 76.53 & 67.46 & 63.72 & 52.63 & 44.53 & 37.86 \\
2.5-3.0 & 0 & 0 & 0 & 0 & 87.10 & 93.50 & 93.14 & 83.35 & 87.40 & 81.77 & 70.88 & 59.00 & 50.68 \\
3.0-3.5 & 0 & 0 & 0 & 105.8 & 68.68 & 89.85 & 99.96 & 100.3 & 95.79 & 96.28 & 84.89 & 68.46 & 61.65 \\
3.5-4.0 & 0 & 0 & 0 & 0 & 56.95 & 87.66 & 88.70 & 102.5 & 99.54 & 101.7 & 91.52 & 76.95 & 71.00 \\
4.0-4.5 & 0 & 0 & 0 & 52.91 & 30.15 & 57.71 & 75.40 & 96.55 & 96.39 & 99.58 & 91.87 & 86.41 & 76.18 \\
4.5-5.0 & 0 & 0 & 0 & 0 & 21.78 & 29.95 & 48.79 & 62.46 & 86.05 & 95.52 & 92.61 & 87.11 & 77.88 \\
5.0-5.5 & 0 & 0 & 0 & 0 & 1.675 & 19.72 & 34.80 & 52.56 & 61.46 & 78.22 & 83.68 & 86.25 & 80.97 \\
5.5-6.0 & 0 & 0 & 0 & 0 & 5.025 & 6.574 & 20.13 & 32.11 & 46.17 & 63.97 & 78.02 & 78.90 & 79.19 \\
6.0-7.0 & 0 & 0 & 0 & 0 & 0 & 7.305 & 12.96 & 29.47 & 46.32 & 78.38 & 116.6 & 134.8 & 144.6 \\
7.0-8.0 & 0 & 0 & 0 & 0 & 0 & 0.7305 & 3.070 & 7.037 & 12.59 & 34.80 & 67.69 & 99.43 & 119.3 \\
8.0-9.0 & 0 & 0 & 0 & 0 & 1.675 & 0 & 0.6823 & 0.6598 & 3.598 & 11.77 & 32.07 & 59.70 & 83.21 \\
9.0-10.0 & 0 & 0 & 0 & 0 & 0 & 0 & 0 & 1.539 & 1.649 & 3.297 & 11.47 & 29.87 & 52.54 \\
10.0-11.0 & 0 & 0 & 0 & 0 & 0 & 0.7305 & 2.047 & 1.759 & 1.499 & 4.464 & 8.036 & 20.14 & 40.87 \\
\hline
\end{tabular}
\caption{$\mu^{-}$ reconstructed from $\nu_{\tau}$ CC events. All values $\times 10^{-4}$.}
\end{table}

\begin{table}[h!] 
\begin{tabular}{|c||c|c|c|c|c|c|c|c|c|c|c|c|c|}
\hline
 & 0.0-2.0 & 2.0-2.5 & 2.5-3.0 & 3.0-3.5 & 3.5-4.0 & 4.0-4.5 & 4.5-5.0 & 5.0-5.5 & 5.5-6.0 & 6.0-7.0 & 7.0-8.0 & 8.0-9.0 & 9.0-10.0 \\
 \hline 
0.0-2.0 & 0 & 0 & 0 & 0 & 0 & 0 & 0 & 0.7401 & 1.547 & 1.304 & 2.603 & 2.494 & 1.652 \\
2.0-2.5 & 0 & 0 & 0 & 0 & 0 & 0 & 0 & 0 & 0.9282 & 1.053 & 1.547 & 1.474 & 2.102 \\
2.5-3.0 & 0 & 0 & 0 & 0 & 0 & 0 & 0.4417 & 0.7401 & 0.3094 & 1.354 & 2.112 & 2.003 & 2.027 \\
3.0-3.5 & 0 & 0 & 0 & 0 & 0 & 0 & 0 & 0.7401 & 0 & 1.104 & 1.282 & 1.625 & 2.628 \\
3.5-4.0 & 0 & 0 & 0 & 0 & 0 & 0 & 0.4417 & 1.234 & 1.083 & 1.154 & 1.584 & 1.965 & 1.577 \\
4.0-4.5 & 0 & 0 & 0 & 0 & 0 & 0 & 0 & 0.7401 & 0.7735 & 1.003 & 1.018 & 1.172 & 2.027 \\
4.5-5.0 & 0 & 0 & 0 & 0 & 0 & 0 & 0 & 0.4934 & 0.4641 & 1.154 & 0.7921 & 1.172 & 1.426 \\
5.0-5.5 & 0 & 0 & 0 & 0 & 0 & 0 & 0.4417 & 0 & 0.4641 & 0.3010 & 1.094 & 1.058 & 0.8258 \\
5.5-6.0 & 0 & 0 & 0 & 0 & 0 & 0 & 0 & 0 & 0 & 0.6019 & 0.4149 & 0.7937 & 0.6006 \\
6.0-7.0 & 0 & 0 & 0 & 0 & 0 & 0 & 0.8833 & 0.2467 & 0.3094 & 0.5518 & 0.8298 & 1.020 & 1.276 \\
7.0-8.0 & 0 & 0 & 0 & 0 & 0 & 0 & 0 & 0 & 0.4641 & 0.2006 & 0.6412 & 0.7559 & 1.051 \\
8.0-9.0 & 0 & 0 & 0 & 0 & 0 & 0 & 0 & 0 & 0 & 0.1505 & 0.1509 & 0.5291 & 0.6006 \\
9.0-10.0 & 0 & 0 & 0 & 0 & 0 & 0 & 0 & 0 & 0.1547 & 0.1505 & 0.1509 & 0.1890 & 0.2252 \\
10.0-11.0 & 0 & 0 & 0 & 0 & 0 & 0 & 0 & 0.4934 & 0 & 0.5016 & 0.6035 & 0.7559 & 0.9009 \\
\hline
\end{tabular}
\caption{$\mu^{-}$ background from $\bar{\nu}_{\tau}$ CC events. All values $\times 10^{-4}$.}
\end{table}

\begin{table}[h!] 
\subsection{$\bar{\nu}_{\mu}$ Appearance Matrices, Negative Focussing Detector Field}
\begin{tabular}{|c||c|c|c|c|c|c|c|c|c|c|c|c|c|}
\hline
 & 0.0-2.0 & 2.0-2.5 & 2.5-3.0 & 3.0-3.5 & 3.5-4.0 & 4.0-4.5 & 4.5-5.0 & 5.0-5.5 & 5.5-6.0 & 6.0-7.0 & 7.0-8.0 & 8.0-9.0 & 9.0-10.0 \\
 \hline 
0.0-2.0 & 1230 & 1305 & 635.2 & 274.4 & 133.1 & 64.81 & 36.28 & 21.88 & 16.17 & 11.30 & 7.222 & 5.422 & 4.818 \\
2.0-2.5 & 569.1 & 1211 & 786.2 & 405.1 & 190.1 & 89.38 & 47.20 & 27.00 & 19.19 & 9.344 & 6.807 & 4.826 & 3.516 \\
2.5-3.0 & 219.1 & 1161 & 1383 & 820.8 & 399.9 & 187.7 & 92.18 & 50.58 & 28.79 & 17.07 & 10.65 & 6.733 & 5.470 \\
3.0-3.5 & 73.04 & 639.8 & 1359 & 1264 & 772.6 & 382.3 & 199.5 & 98.06 & 57.71 & 27.31 & 14.19 & 11.98 & 8.074 \\
3.5-4.0 & 18.69 & 234.1 & 847.2 & 1387 & 1202 & 724.0 & 381.1 & 193.5 & 97.01 & 46.16 & 22.39 & 14.24 & 10.29 \\
4.0-4.5 & 11.89 & 104.2 & 436.0 & 1037 & 1343 & 1142 & 679.5 & 363.1 & 190.7 & 88.80 & 34.66 & 19.90 & 13.41 \\
4.5-5.0 & 8.493 & 20.84 & 138.2 & 569.0 & 1106 & 1290 & 1048 & 663.3 & 366.2 & 156.1 & 57.78 & 30.09 & 20.32 \\
5.0-5.5 & 5.096 & 22.06 & 46.79 & 221.4 & 641.7 & 1127 & 1220 & 989.9 & 635.2 & 279.4 & 98.00 & 42.72 & 28.13 \\
5.5-6.0 & 8.493 & 12.26 & 25.52 & 80.10 & 291.6 & 735.5 & 1108 & 1163 & 919.3 & 475.4 & 164.1 & 68.94 & 35.55 \\
6.0-7.0 & 8.493 & 22.06 & 26.94 & 45.58 & 157.5 & 514.8 & 1207 & 1852 & 2119 & 1602 & 735.6 & 294.9 & 125.9 \\
7.0-8.0 & 5.096 & 12.26 & 14.18 & 15.19 & 32.88 & 92.03 & 280.6 & 726.8 & 1336 & 1852 & 1458 & 729.8 & 342.0 \\
8.0-9.0 & 16.99 & 11.03 & 9.925 & 14.27 & 17.54 & 23.61 & 62.01 & 148.2 & 417.6 & 1141 & 1709 & 1343 & 762.2 \\
9.0-10.0 & 3.397 & 7.354 & 6.380 & 4.604 & 13.15 & 15.18 & 24.25 & 33.36 & 90.44 & 415.0 & 1175 & 1550 & 1279 \\
10.0-11.0 & 5.096 & 7.354 & 11.34 & 14.27 & 16.60 & 19.76 & 27.02 & 41.43 & 57.71 & 173.0 & 726.4 & 2023 & 3398 \\
\hline
\end{tabular}
\caption{Golden channel $\bar{\nu}_{\mu}$ appearance signal efficiency. All values $\times 10^{-4}$.}
\end{table}

\begin{table}[h!] 
\begin{tabular}{|c||c|c|c|c|c|c|c|c|c|c|c|c|c|}
\hline
 & 0.0-2.0 & 2.0-2.5 & 2.5-3.0 & 3.0-3.5 & 3.5-4.0 & 4.0-4.5 & 4.5-5.0 & 5.0-5.5 & 5.5-6.0 & 6.0-7.0 & 7.0-8.0 & 8.0-9.0 & 9.0-10.0 \\
 \hline 
0.0-2.0 & 2.052 & 6.321 & 1.787 & 3.720 & 1.767 & 1.651 & 2.380 & 1.724 & 1.579 & 1.932 & 0.7482 & 0.9795 & 0.6611 \\
2.0-2.5 & 1.368 & 0 & 1.191 & 0.4134 & 0.5889 & 0.4718 & 0.1831 & 0.3134 & 0.6578 & 0.2273 & 0.1603 & 0.4285 & 0.2644 \\
2.5-3.0 & 0 & 0 & 0 & 0.4134 & 0.8833 & 0.7077 & 0.7323 & 0.7836 & 0.6578 & 0.8525 & 0.6413 & 0.4285 & 0.6611 \\
3.0-3.5 & 0 & 0 & 1.787 & 0.8268 & 1.767 & 0.9436 & 1.465 & 0.4701 & 0.5262 & 0.4546 & 0.7482 & 0.1836 & 0.6611 \\
3.5-4.0 & 0.6839 & 0 & 1.191 & 1.240 & 1.178 & 0.9436 & 0.7323 & 0.7836 & 0.2631 & 0.7388 & 0.6948 & 0.5509 & 1.190 \\
4.0-4.5 & 0 & 0.9030 & 0.5956 & 2.894 & 1.178 & 0.4718 & 0.9153 & 0.9403 & 0.9209 & 0.8525 & 0.8551 & 0.3673 & 0.3966 \\
4.5-5.0 & 0 & 0 & 0 & 1.654 & 2.944 & 0.7077 & 1.281 & 1.097 & 1.184 & 0.8525 & 0.6948 & 0.4897 & 0.2644 \\
5.0-5.5 & 1.368 & 0 & 1.191 & 2.067 & 1.178 & 1.651 & 0.7323 & 1.254 & 0.9209 & 1.307 & 1.229 & 0.5509 & 0.7933 \\
5.5-6.0 & 0 & 0 & 0.5956 & 0.4134 & 0.2944 & 0.9436 & 1.281 & 0.9403 & 1.579 & 1.762 & 1.015 & 0.9182 & 0.3966 \\
6.0-7.0 & 0 & 0.9030 & 1.191 & 2.067 & 1.178 & 3.067 & 3.478 & 2.664 & 2.894 & 2.103 & 2.245 & 1.898 & 1.587 \\
7.0-8.0 & 0 & 0 & 1.191 & 1.654 & 1.472 & 1.887 & 2.746 & 2.507 & 3.289 & 2.785 & 2.458 & 2.510 & 0.7933 \\
8.0-9.0 & 0.6839 & 1.806 & 0 & 0 & 0.5889 & 0.9436 & 1.098 & 1.567 & 2.500 & 2.273 & 2.512 & 1.898 & 1.322 \\
9.0-10.0 & 0 & 0 & 0 & 0.8268 & 0.5889 & 0 & 0.7323 & 1.724 & 1.052 & 1.364 & 2.031 & 2.143 & 2.776 \\
10.0-11.0 & 2.052 & 0 & 0 & 0.8268 & 0.8833 & 1.415 & 2.014 & 1.881 & 3.815 & 4.944 & 5.986 & 7.652 & 10.44 \\
\hline
\end{tabular}
\caption{$\mu^{+}$ background from charge mis-identified $\nu_{\mu}$ CC events All values $\times 10^{-4}$.}
\end{table}

\begin{table}[h!] 
\begin{tabular}{|c||c|c|c|c|c|c|c|c|c|c|c|c|c|}
\hline
 & 0.0-2.0 & 2.0-2.5 & 2.5-3.0 & 3.0-3.5 & 3.5-4.0 & 4.0-4.5 & 4.5-5.0 & 5.0-5.5 & 5.5-6.0 & 6.0-7.0 & 7.0-8.0 & 8.0-9.0 & 9.0-10.0 \\
 \hline 
0.0-2.0 & 0 & 0 & 0 & 0 & 64.21 & 67.39 & 63.16 & 66.61 & 63.11 & 58.04 & 57.26 & 47.66 & 48.50 \\
2.0-2.5 & 0 & 0 & 0 & 344.8 & 21.40 & 54.91 & 68.46 & 60.94 & 57.55 & 57.13 & 50.32 & 46.71 & 43.69 \\
2.5-3.0 & 0 & 0 & 0 & 0 & 80.26 & 76.13 & 79.06 & 83.63 & 77.04 & 77.35 & 73.48 & 63.00 & 57.73 \\
3.0-3.5 & 0 & 0 & 0 & 0 & 58.86 & 92.35 & 95.84 & 99.18 & 90.03 & 91.04 & 87.66 & 74.53 & 63.51 \\
3.5-4.0 & 0 & 0 & 0 & 0 & 69.56 & 92.35 & 85.68 & 93.75 & 93.13 & 92.85 & 89.21 & 83.71 & 72.90 \\
4.0-4.5 & 0 & 0 & 0 & 0 & 53.50 & 41.18 & 64.92 & 73.52 & 83.69 & 89.59 & 92.04 & 86.32 & 79.73 \\
4.5-5.0 & 0 & 0 & 0 & 0 & 16.05 & 37.44 & 49.02 & 65.13 & 72.40 & 81.31 & 84.11 & 81.97 & 77.63 \\
5.0-5.5 & 0 & 0 & 0 & 0 & 16.05 & 17.47 & 29.59 & 43.91 & 53.99 & 68.07 & 75.40 & 76.42 & 78.75 \\
5.5-6.0 & 0 & 0 & 0 & 0 & 5.350 & 6.240 & 15.46 & 26.64 & 33.57 & 51.67 & 68.42 & 70.03 & 74.85 \\
6.0-7.0 & 0 & 0 & 0 & 0 & 0 & 4.992 & 11.48 & 21.96 & 33.57 & 65.56 & 95.81 & 115.5 & 121.2 \\
7.0-8.0 & 0 & 0 & 0 & 0 & 0 & 0 & 1.767 & 6.414 & 12.22 & 26.43 & 55.64 & 76.87 & 97.97 \\
8.0-9.0 & 0 & 0 & 0 & 0 & 0 & 1.248 & 0 & 0.4934 & 2.784 & 8.277 & 23.54 & 45.31 & 65.46 \\
9.0-10.0 & 0 & 0 & 0 & 0 & 0 & 0 & 0.8833 & 0.4934 & 1.702 & 2.157 & 7.091 & 21.24 & 34.83 \\
10.0-11.0 & 0 & 0 & 0 & 0 & 5.350 & 2.496 & 2.208 & 1.234 & 1.856 & 3.010 & 4.677 & 12.62 & 30.40 \\
\hline
\end{tabular}
\caption{$\mu^{+}$ reconstructed from $\bar{\nu}_{\tau}$ CC events. All values $\times 10^{-4}$.}
\end{table}

\begin{table}[h!] 
\begin{tabular}{|c||c|c|c|c|c|c|c|c|c|c|c|c|c|}
\hline
 & 0.0-2.0 & 2.0-2.5 & 2.5-3.0 & 3.0-3.5 & 3.5-4.0 & 4.0-4.5 & 4.5-5.0 & 5.0-5.5 & 5.5-6.0 & 6.0-7.0 & 7.0-8.0 & 8.0-9.0 & 9.0-10.0 \\
 \hline 
0.0-2.0 & 0 & 0 & 0 & 0 & 0 & 0.7305 & 1.706 & 0.8797 & 2.848 & 2.892 & 3.277 & 3.753 & 4.481 \\
2.0-2.5 & 0 & 0 & 0 & 0 & 0 & 0.7305 & 0.6823 & 1.539 & 1.049 & 2.486 & 1.756 & 2.737 & 2.627 \\
2.5-3.0 & 0 & 0 & 0 & 0 & 0 & 0 & 1.023 & 0.8797 & 1.199 & 1.420 & 2.341 & 3.089 & 3.090 \\
3.0-3.5 & 0 & 0 & 0 & 0 & 0 & 0.7305 & 0.3412 & 1.100 & 1.199 & 1.826 & 2.029 & 2.815 & 3.322 \\
3.5-4.0 & 0 & 0 & 0 & 0 & 0 & 0 & 1.706 & 1.759 & 0.1499 & 1.623 & 1.756 & 2.463 & 2.318 \\
4.0-4.5 & 0 & 0 & 0 & 0 & 0 & 1.461 & 0.3412 & 1.320 & 1.199 & 1.928 & 1.599 & 1.955 & 2.627 \\
4.5-5.0 & 0 & 0 & 0 & 0 & 0 & 0 & 0 & 0.4398 & 0.4497 & 1.015 & 1.170 & 2.189 & 1.700 \\
5.0-5.5 & 0 & 0 & 0 & 0 & 0 & 0.7305 & 0.3412 & 0.4398 & 0.2998 & 0.7102 & 1.092 & 1.251 & 2.859 \\
5.5-6.0 & 0 & 0 & 0 & 0 & 0 & 0.7305 & 0.3412 & 0.4398 & 0.1499 & 0.7102 & 1.053 & 1.447 & 1.004 \\
6.0-7.0 & 0 & 0 & 0 & 0 & 0 & 0.7305 & 0 & 0.4398 & 0.7495 & 0.7102 & 1.678 & 2.502 & 2.472 \\
7.0-8.0 & 0 & 0 & 0 & 0 & 0 & 0 & 0 & 0.2199 & 0 & 0.6087 & 1.014 & 1.603 & 1.700 \\
8.0-9.0 & 0 & 0 & 0 & 0 & 0 & 0 & 0.3412 & 0.4398 & 0.1499 & 0.2029 & 0.2341 & 0.7429 & 1.082 \\
9.0-10.0 & 0 & 0 & 0 & 0 & 0 & 0 & 0 & 0.2199 & 0 & 0.3044 & 0.3901 & 0.3910 & 0.6953 \\
10.0-11.0 & 0 & 0 & 0 & 0 & 1.675 & 0 & 0 & 0 & 0.1499 & 0.4566 & 0.6632 & 0.9775 & 1.468 \\
\hline
\end{tabular}
\caption{$\mu^{+}$ background from $\nu_{\tau}$ CC events. All values $\times 10^{-4}$.}
\end{table}

\end{document}